\documentclass[%
 reprint,
 superscriptaddress,
nofootinbib,
 amsmath,amssymb,
 aps,
floatfix,
]{revtex4-2}

\usepackage{graphicx}
\usepackage{dcolumn}
\usepackage{bm}
\usepackage[pdftex,
            pdftitle={Fermion Discretization Effects in the Two-Flavor Lattice Schwinger Model: A Study with Matrix Product States},
            pdfauthor={Tim Schwaegerl, Karl Jansen, Stefan Kuehn},
            bookmarks,
            colorlinks,          
            menucolor=black,            
            plainpages=false,
            pdfpagelabels,
            hypertexnames=false]{hyperref}
\usepackage{booktabs}  
\usepackage{multirow}  
\usepackage[english]{babel} 
\usepackage{xcolor}  

\definecolor{desyblue}{RGB}{0, 166, 235}

\begin{document}

\title{
Fermion Discretization Effects in the Two-Flavor Lattice Schwinger Model: \\A Study with Matrix Product States}

\author{Tim Schw\"agerl}
\affiliation{Deutsches Elektronen-Synchrotron DESY, Platanenallee 6, 15738 Zeuthen, Germany}
\affiliation{Institut f\"ur Physik, Humboldt-Universit\"at zu Berlin, Newtonstr. 15, 12489 Berlin, Germany}

\author{Karl Jansen}
\affiliation{Computation-Based Science and Technology Research Center, The Cyprus Institute, 20 Kavafi Street, 2121 Nicosia, Cyprus}
\affiliation{Deutsches Elektronen-Synchrotron DESY, Platanenallee 6, 15738 Zeuthen, Germany}

\author{Stefan K\"uhn}
\affiliation{Deutsches Elektronen-Synchrotron DESY, Platanenallee 6, 15738 Zeuthen, Germany}

\date{\today}

\begin{abstract}
We present a comprehensive tensor network study of staggered, Wilson, and twisted mass fermions in the Hamiltonian formulation, using the massive two-flavor Schwinger model as a benchmark. Particular emphasis is placed on twisted mass fermions, whose properties in this context have not been systematically explored before. 
We confirm the expected \( \mathcal{O}(a) \) improvement in the free theory and observe that this improvement persists in the interacting case. By leveraging an electric-field-based method for mass renormalization, we reliably tune to maximal twist and establish the method’s applicability in the two-flavor model. Once mass renormalization is included, the pion mass exhibits rapid convergence to the continuum limit. Finite-volume effects are addressed using two complementary approaches: dispersion relation fits and finite-volume scaling. 
Our results show excellent agreement with semiclassical predictions and reveal a milder volume dependence for twisted mass fermions compared to staggered and Wilson discretizations. In addition, we observe clear isospin-breaking effects, suggesting intriguing parallels with lattice QCD. These findings highlight the advantages of twisted mass fermions for Hamiltonian simulations and motivate their further exploration—particularly in view of future applications to higher-dimensional lattice gauge theories.

\end{abstract}

\maketitle

\section{\label{sec:intro}Introduction}
Lattice gauge theory (LGT) provides a powerful nonperturbative framework for studying gauge theories~\cite{Gattringer2010, Rothe2012}. By discretizing the Lagrangian on a Euclidean space-time lattice, it becomes amenable to numerical treatment via Markov Chain Monte Carlo (MC) methods. These techniques have enabled high-precision studies of hadronic physics in Quantum Chromodynamics (QCD)~\cite{FLAG2024}. However, standard MC methods suffer from the infamous sign problem in certain regimes, which severely restricts their applicability. Notably, QCD at finite baryon chemical potential or in the presence of a topological \(\theta\)-term remains largely inaccessible, as do real-time dynamical phenomena that require a formulation in Minkowski space-time~\cite{Loh1990}.

The Hamiltonian formulation of LGTs~\cite{Kogut1975} offers a promising alternative. Numerical methods based on this formulation typically do not rely on sampling configurations, thus avoiding the sign problem that hinders MC techniques in certain regimes. In particular, numerical methods based on tensor network states (TNS), which provide an efficient representation of moderately entanglement quantum states, have proven themselves as a powerful tool for accessing LGTs in sign-problem afflicted regimes. Tensor network approaches have achieved impressive success for \(1+1\) dimensional models~\cite{Banuls2019,Banuls2019a,Banuls2020,Rigobello2021,Rigobello2023,Papaefstathiou2024,Angelides2025a,Chai2025}. While generalizing this success to higher dimensions is not an immediate task, recent years have shown promising developments and the first simulations with TNS in higher dimensions have been carried out~\cite{Felser2019, Magnifico2021, Cataldi2024, Magnifico2024}.

A crucial aspect of lattice formulations is the treatment of the fermionic degrees of freedom. Naively discretized fermion fields lead to redundant species when taking the continuum limit--a phenomenon know as fermion doubling--necessitating specialized discretization schemes to recover the correct continuum theory~\cite{Gattringer2010}. Although a wide range of fermion discretization schemes exists~\cite{Kogut1975,Ginsparg1982,Aoki1984,Luescher1998,Frezzotti2001,Karsten1981,Wilczek1987,Borici2008,Quinn1986,Weber2015}, each comes with its own set of advantages and limitations.

In the context of TNS, the majority of existing work has focused on staggered fermions~\cite{Banuls2019,Banuls2019a,Banuls2020}. This formulation effectively thins out the fermionic degrees of freedom, making it computationally efficient. However, it does not fully eliminate fermion doublers beyond \(1+1\) dimensions, limiting its suitability for computations in higher dimensions that aim to take the continuum limit. Recently, the first TNS simulations using Wilson fermions have been carried out for the lattice Schwinger model~\cite{Angelides2023}, introducing a procedure for computing the additive mass renormalization in the Hamiltonian formulation. While this discretization fully removes the doublers in arbitrary dimensions, it is generally computationally more expensive than the staggered formulation. In addition, it explicitly breaks chiral symmetry, resulting in a larger additive mass renormalization.

A very important aspect of LGT simulations is the approach to the continuum limit, achieved by taking the lattice spacing \(a\) to zero. Staggered fermions are expected to scale with \(\mathcal{O}(a^2)\) corrections, whereas Wilson fermions exhibit linear \(\mathcal{O}(a)\) scaling. To suppress these leading-order discretization effects, the Symanzik improvement programme was developed and successfully applied in LGT simulations. This approach involves improving both the lattice action and relevant observables, leading to \(\mathcal{O}(a^2)\) scaling for Symanzik-improved Wilson fermions. However, this procedure is technically demanding and must be tailored to each observable individually.

As an alternative, twisted mass fermions~\cite{Shindler2008} at maximal twist provide automatic \(\mathcal{O}(a)\) improvement without requiring the full Symanzik programme. This makes them an attractive formulation for precision studies. Despite their successful use in the Lagrangian formalism, twisted mass fermions have not yet been systematically explored in the Hamiltonian framework. While they ensure faster convergence to the continuum limit at maximal twist, they also introduce explicit isospin breaking at finite lattice spacing, resulting in characteristic effects such as pion mass splitting. Related isospin-breaking effects due to nondegenerate quark masses have also been studied recently in Ref.~\cite{Albandea2025}.

In this work, we study three representative fermion discretizations, staggered fermions~\cite{Kogut1975}, Wilson fermions~\cite{Wilson1974}, and twisted mass Wilson fermions~\cite{Aoki1984} using the two-flavor Schwinger model as a benchmark. This \(1+1\) dimensional Abelian gauge theory shares important qualitative features with QCD, such as confinement and a nontrivial vacuum structure, while remaining computationally accessible. The one-flavor Schwinger model has been extensively studied using TNS methods~\cite{Banuls2016, Funcke2020, Rigobello2021, Zache2022}, and recent work has extended these studies to the two-flavor model~\cite{Banuls2017, Itou2023, Funcke2023, Itou2024, Dempsey2024}, which exhibits a rich vacuum structure and excitation spectrum.
Moreover, the lattice Schwinger model has recently attracted significant attention as a testbed for quantum simulations~\cite{Kuehn2022, Farrell2024, Bringewatt2024, Angelides2025, Schuster2024}, including applications such as the computation of excited states~\cite{guo2024}. The two-flavor model was also the subject of early scaling studies of various fermion discretizations—including twisted mass fermions—within the Lagrangian framework~\cite{Christian2006}.

The additive mass renormalization in the Schwinger model has received renewed attention in recent years. An analytical prediction was derived for staggered fermions with periodic boundary conditions~\cite{Dempsey2022}, and its inclusion was shown to significantly improve continuum convergence. A subsequent study introduced a numerical method based on the electric field density within the Hamiltonian formalism, successfully applying it to Wilson fermions in the one-flavor case~\cite{Angelides2023}. Here, we extend this method to the two-flavor model, which presents additional challenges due to a gapless phase near the region of interest. We demonstrate that the method remains applicable, even in this regime. This allows us to investigate the continuum-limit behavior for all three fermion discretizations with high accuracy and to test the expected scaling with the lattice spacing.

We then turn our attention to the low-energy spectrum, focusing on the extraction of the pion mass. In particular, twisted mass fermions are known to exhibit isospin breaking at finite lattice spacing, a phenomenon familiar from lattice QCD~\cite{Shindler2008}. To correct for finite-volume effects and obtain precise estimates of the pion mass, we employ two complementary techniques: finite-volume scaling~\cite{Banuls2013} and dispersion relation fits~\cite{Itou2023, Itou2024}. The resulting values are compared against analytical predictions~\cite{Smilga1997, Hosotani1998} and serve to investigate discretization effects across the different fermion formulations.

This paper is structured as follows. In Sec.~\ref{sec:schwinger_model}, we introduce the massive two-flavor Schwinger model and summarize relevant analytical predictions. In Sec.~\ref{sec:discretizations}, we review the staggered and Wilson fermion discretizations. We then move on to twisted mass fermions in Sec.~\ref{sec:twisted_mass}, beginning with their formulation in the continuum, followed by a discussion of their lattice implementation and the mechanism of automatic \( \mathcal{O}(a) \) improvement. 
After these theoretical considerations, we describe the setup of our TNS simulations in Sec.~\ref{sec:mps}, and present our numerical results in Sec.~\ref{sec:results}. We begin with the scaling behavior of the electric field density and demonstrate automatic \( \mathcal{O}(a) \) improvement in Sec.~\ref{sec:electric_field_density}. The determination of the mass renormalization and its influence on scaling is presented in Sec.~\ref{sec:mass_renormalization_results}. We then analyze the low-energy spectrum for the different discretizations in Sec.~\ref{sec:mass_spectrum}. To compute the pion mass, we account for finite volume effects via the dispersion relation in Sec.~\ref{sec:pion_mass_dispersion} and via finite-volume scaling in Sec.~\ref{sec:pion_finite_size}. We conclude with a summary and outlook in Sec.~\ref{sec:discussion}.

\section{\label{sec:schwinger_model}The massive two-flavor Schwinger model}
The massive two-flavor Schwinger model describes quantum electrodynamics in \((1+1)\)-dimensions coupled to massive Dirac fermions~\cite{Schwinger1962}. The Lagrangian density is given by
\begin{equation}
    \begin{split}
    \mathcal{L} &= \underbrace{\sum_{f=0}^{1} \left( i\bar{\psi}_f \gamma^{\mu} D_{\mu} \psi_f - m \bar{\psi}_f \psi_f \right)}_{\mathcal{L}_F} \\
    &\quad \underbrace{- \frac{1}{4} F_{\mu\nu} F^{\mu\nu} + \frac{g\theta}{4\pi} \epsilon^{\mu\nu} F_{\mu\nu}}_{\mathcal{L}_G},
    \end{split}
\end{equation}
where the Lagrangian consists of a fermionic part \( \mathcal{L}_F \) and a gauge field part \( \mathcal{L}_G \). In the expression above, the covariant derivative is defined as \( D_{\mu} = \partial_{\mu} + ig A_{\mu} \) with the gauge field \(A_{\mu}\) and the coupling constant \(g\), mediates the interaction between the matter fields. The tensor \(F_{\mu\nu}=\partial_{\mu}A_{\nu}-\partial_{\nu}A_{\mu}\) corresponds to the field strength. The topological \(\theta\)-term, with \(\theta \in [0,2\pi]\), describes the background electric flux and explicitly violates parity.  The fermionic fields \( \psi_f \) are two-component Dirac spinors, and we assume equal masses \( m \) for the all flavors. The model with equal masses exhibits an \( SU(2) \) isospin symmetry between the flavors, with symmetry generators given by
\begin{equation}
\label{eq:isospin}
    J_i = \frac{1}{2} \int dx\, \bar{\psi} \gamma^0 \tau^i \psi,
\end{equation}
where \( \tau^i \) are the Pauli matrices acting in flavor space on the flavor doublet \( \psi \).

The Hamiltonian density of the massive Schwinger model in the temporal gauge, \(A_0=0\), is given by
\begin{equation}
    \begin{split}
    \mathcal{H}&=\sum_{f=0}^{1}\left(-i\bar{\psi}_f \gamma^{1}(\partial_{1}-igA_{1})\psi_f +m\bar{\psi}_f\psi_f \right)\\
    &+ \frac{1}{2}\left(\dot{A}_1 +\frac{g\theta}{2\pi} \right)^2.
    \end{split}
\end{equation}  
Physically relevant states must satisfy Gauss's law,  
\begin{equation}
    -\partial_1\dot{A}^1=g\bar{\psi}\gamma^0\psi.
\end{equation}  
This equation fixes the electric field \(-\dot{A}^1\) up to an integration constant \(g\theta/2\pi\), which corresponds to a homogenous background electric field.

The massless one-flavor Schwinger model is exactly solvable through bosonization, and for small fermion masses, relevant observables can be studied using mass perturbation theory~\cite{Adam1997}. In contrast, the two-flavor Schwinger model exhibits a more intricate structure. In the chiral limit (\( m = 0 \)), the model possesses an \( SU(2) \times SU(2) \) chiral symmetry. At low energies, it reduces to the \( SU(2) \) Wess-Zumino-Witten (WZW) model, a conformal field theory~\cite{Gepner1985,Affleck1986}.

When fermion masses are introduced, the system becomes gapped and develops a non-degenerate ground state for \( \theta \neq \pi \). Unlike the one-flavor case, the analytical study of the two-flavor model requires additional approximations. In the strong coupling regime and for small fermion masses \( m/g \ll 1 \), Ref.~\cite{Smilga1997} derives the following dependence of the mass gap on the fermion mass using effective field theory and scaling arguments. The prefactor is fixed by matching to known universal amplitude ratios, previously computed using the thermodynamic Bethe ansatz in integrable models:
\begin{equation}
    \begin{split}
        \frac{m_{\pi}}{g} &\approx 2^{5/6}e^{\gamma/3}\left(\frac{\Gamma(3/4)}{\Gamma(1/4)}\right)^{2/3}\frac{\Gamma(1/6)}{\Gamma(2/3)}\left(\frac{m}{g}\right)^{2/3}\\
        &\approx 2.008\left(\frac{m}{g}\right)^{2/3},
    \end{split}
\end{equation}
where \( \gamma \approx 0.577 \) is the Euler-Mascheroni constant, and \( \Gamma(\cdot) \) denotes the Gamma function.

In contrast, a semiclassical analysis based on a generalized Hartree-Fock approximation by Ref.~\cite{Hosotani1998}, also valid in the small mass regime \( m/g \ll 1 \), yields
\begin{equation}
\begin{split}
     \frac{m_{\pi}}{g} &\approx e^{2\gamma/3}\frac{2^{5/6}}{\pi^{1/6}} \left(\frac{m}{g}\right)^{2/3}\\
     &\approx 2.163\left(\frac{m}{g}\right)^{2/3}.
\end{split}
\end{equation}

These two theoretical predictions differ by approximately 7\% and are valid only in the regime of small fermion masses, \( m/g \ll 1 \). This relatively small discrepancy necessitates accurate simulations with all systematic errors under control, making the massive two-flavor Schwinger model a compelling target for numerical studies. Such simulations can be used to assess the validity and precision of these analytical approximations.

Additionally, the presence of the topological \( \theta \)-term modifies the overall mass scale, leading to the following dependence of the mass gap on \( \theta \)~\cite{Coleman1976}:
\begin{equation}
     \frac{m_{\pi}}{g} \sim\left|\frac{m}{g}\cos\frac{\theta}{2}\right|^{2/3}.
\end{equation}

\section{\label{sec:discretizations}Lattice discretizations}
To enable numerical simulations within the Hamiltonian formalism, the spatial dimension of the theory must be discretized on a lattice. However, a naive discretization of fermions introduces additional, unphysical fermion species. In \( d \) spatial dimensions, this so-called fermion doubling problem leads to \( 2^d \) fermion species~\cite{Rothe2012, Gattringer2010}. Two widely used strategies to resolve this issue are the staggered and Wilson formulations. In this work, we summarize the corresponding Hamiltonians and define the relevant observables for both approaches.

\subsection{\label{sec:staggered}Staggered fermions}
One approach to discretizing the model on a lattice with an even number of sites \( N \), lattice spacing \( a \), and open boundary conditions is the Kogut-Susskind Hamiltonian~\cite{Kogut1975}:
\begin{equation}
    \begin{split}
    H_S &= -\frac{i}{2a} \sum_{n=0}^{N-2} \sum_{f=0}^{1} (\phi_{n,f}^{\dag} U_n \phi_{n+1,f} - \text{h.c.}) \\
    &\quad + m_{\mathrm{lat}} \sum_{n=0}^{N-1} \sum_{f=0}^{1} (-1)^n \phi_{n,f}^{\dag} \phi_{n,f} + \frac{ag^2}{2} \sum_{n=0}^{N-2} L_n^2.
    \end{split}
\end{equation}
The central idea behind this formulation is to distribute the spinor components across neighboring lattice sites. Here, \( \phi_{n,f} \) denotes a single-component fermionic field of flavor \( f \) at site $n$. In the continuum limit \( a \to 0 \), the fields on even sites correspond to the upper spinor components, and those on odd sites to the lower ones.

The operator \( L_n \) acts on the link between sites \( n \) and \( n+1 \), representing the dimensionless, quantized electric flux. Its conjugate operator \( U_n \), satisfying \( [U_n, L_{n'}] = \delta_{nn'} U_n \), acts as a lowering operator for the electric flux. Gauss's law on the lattice reads
\begin{equation}
    L_n - L_{n-1} = Q_n,
\end{equation}
with the staggered charge operator defined as
\begin{equation}
    Q_n = \sum_{f=0}^{1} \phi_{n,f}^{\dag} \phi_{n,f} - \frac{F}{2}(1 - (-1)^n).
\end{equation}
For open boundary conditions, fixing \( L_0 = l_0 \) allows for a recursive solution:
\begin{equation}
    L_n = l_0 + \sum_{k=0}^n Q_k.
\end{equation}
Here, the background field \( l_0 = \theta / (2\pi) \) corresponds to the topological \( \theta \)-term in the continuum limit.

To facilitate numerical simulations, we rescale the Hamiltonian to a dimensionless form \( W = (2/ag^2) H \). After applying a residual gauge transformation~\cite{Hamer1997a}, the rescaled staggered Hamiltonian becomes
\begin{equation}
\begin{split}
    W_S &= -ix \sum_{n=0}^{N-2} \sum_{f=0}^{1} (\phi_{n,f}^{\dag} \phi_{n+1,f} - \text{h.c.}) \\
    &\quad + 2\sqrt{x} \frac{m_{\mathrm{lat}}}{g} \sum_{n=0}^{N-1} \sum_{f=0}^{1} (-1)^n \phi_{n,f}^{\dag} \phi_{n,f} \\
    &\quad + \sum_{n=0}^{N-2} \left( l_0 + \sum_{k=1}^n Q_k \right)^2,
\end{split}
\end{equation}
where \( x = 1/(ag)^2 \) denotes the inverse squared lattice spacing in units of the coupling.

Our primary observable is the electric field density. To mitigate staggering artifacts and boundary effects, we define it as the average over two adjacent links near the lattice center:
\begin{equation}
    L_S = l_0 + \frac{1}{2} \left( \sum_{n=0}^{N/2 - 2} Q_n + \sum_{n=0}^{N/2 - 1} Q_n \right).
\end{equation}

In order to enable the extraction of dispersion relations, we start from the continuum expression for the momentum operator~\cite{Rothe2012}: 
\begin{equation}
\label{eq:momentum_continuum}
    P = \sum_{f=0}^{1} \int dx\, \psi_f^{\dag} i\partial_x\psi_f.
\end{equation}
We then adopt a standard symmetric discretization for the momentum operator on the lattice~\cite{Banuls2013, Itou2023}:
\begin{equation}
\label{eq:momentum_staggered}
    \frac{K}{g} = \frac{i}{4}\sqrt{x} \sum_{n=0}^{N-3} \sum_{f=0}^{1} \left( \phi_{n,f}^{\dag} \phi_{n+2,f} - \text{h.c.} \right),
\end{equation}
where the prefactor ensures proper normalization in terms of the dimensionless coupling parameter \( x = 1/(ag)^2 \). This pseudo-momentum operator does not exactly commute with the Hamiltonian due to the use of open boundary conditions, which break translational invariance. However, it still provides a reliable estimate for the momentum and can be used to approximately characterize excitation energies as a function of momentum.

Similarly, a discretization of the continuum isospin generators defined in Eq.~\eqref{eq:isospin} yields corresponding lattice operators \( J_i \) for the three isospin components. The total isospin is then given by \( J^2 = J_x^2 + J_y^2 + J_z^2 \).

Another key observable is the excitation gap, defined as
\begin{equation}
    \frac{\Delta E}{g} = \frac{W_1 - W_0}{2\sqrt{x}},
\end{equation}
where \( W_0 \) and \( W_1 \) denote the ground state and first excited state energies of the rescaled Hamiltonian \( W \), respectively. The division by \(2\sqrt{x}\) converts the dimensionless energy difference obtained from the numerical simulations into a physical quantity with dimensions of energy, expressed in units of the coupling \(g\).

In one spatial dimension (\( d = 1 \)), staggered fermions completely eliminate the doubling problem. However, in higher dimensions, they leave behind residual degrees of freedom, commonly referred to as tastes~\cite{Rothe2012}. A formulation that removes all fermion doublers in arbitrary dimensions is provided by Wilson fermions, which we discuss in the next section.

\subsection{\label{sec:wilson}Wilson fermions}
The Wilson fermion formulation resolves the fermion doubling problem by introducing an additional second-derivative term—known as the Wilson term—into the Hamiltonian~\cite{Wilson1974}. This term acts as a momentum-dependent mass that suppresses the unwanted fermion doublers by giving them masses of order \(1/a\), where \(a\) is the lattice spacing. While this term explicitly breaks chiral symmetry at finite lattice spacing, it vanishes smoothly in the continuum limit, restoring the correct symmetries of the theory. Intuitively, the Wilson term modifies the fermion dispersion relation so that only one low-energy mode remains physical, while the spurious modes acquire large masses and decouple from the low-energy physics as \(a \to 0\). This approach trades exact chiral symmetry for the ability to control fermion doubling in lattice simulations.

The Wilson formulation has also been argued to offer advantages over staggered fermions in the context of simulating the Schwinger model with ultracold atoms in optical lattices~\cite{Zache2018}. Building on the approach of Refs.~\cite{Zache2018, Mazzola2021, Angelides2023}, we generalize their single-flavor results to construct the Hamiltonian for the two-flavor Schwinger model:

\begin{equation}
    \begin{split}
    H_W &=\sum_{n=0}^{N-2}\sum_{f=0}^{1}\left(\bar{\phi}_{n,f}\left(\frac{1+i\gamma_1}{2a}\right)U_n\phi_{n+1,f} + \text{h.c.} \right) \\
    &\quad + \left(m_{\mathrm{lat}} + \frac{1}{a}\right) \sum_{n=0}^{N-1} \sum_{f=0}^{1} \bar{\phi}_{n,f} \phi_{n,f} \\
    &\quad + \frac{ag^2}{2} \sum_{n=0}^{N-2} L_n^2.
    \end{split}
\end{equation}

Unlike the staggered formulation, \( \phi_{n,f} \) denotes a dimensionless two-component Dirac spinor at site \( n \) with flavor index \( f \). We choose the convention \( \gamma^0 = X \), \( \gamma^1 = iZ \), where \( X \) and \( Z \) are Pauli matrices. The local charge operator is given by
\begin{equation}
    Q_n = \sum_{f=0}^{1} \left( \phi_{n,f,1}^\dag \phi_{n,f,1} + \phi_{n,f,2}^\dag \phi_{n,f,2} \right) - F.
\end{equation}

As in the staggered case, the gauge fields can be integrated out using Gauss's law. Applying this and rescaling to a dimensionless Hamiltonian \( W = (2/ag^2) H \), we obtain:
\begin{equation}
\begin{split}
    W_W &= 
    2\sum_{n=0}^{N-1} \sum_{f=0}^{1} \left( \frac{m_{\mathrm{lat}}}{g} \sqrt{x} + x \right) \left( \phi_{n,f,1}^\dag \phi_{n,f,2} + \text{h.c.} \right) \\
    &\quad + 2x \sum_{n=0}^{N-2} \sum_{f=0}^{1} \left( \phi_{n,f,1}^\dag \phi_{n+1,f,2} + \text{h.c.} \right) \\
    &\quad + \sum_{n=0}^{N-2} \left( l_0 + \sum_{k=1}^n Q_k \right)^2,
\end{split}
\end{equation}
where \( x = 1/(ag)^2 \).

The electric field density becomes approximately homogeneous for sufficiently large systems. To suppress boundary effects, we evaluate it at the center of the chain using
\begin{equation}
    L_W = l_0 + \frac{1}{2} \sum_{n=0}^{\lceil N/2 \rceil - 1} Q_n.
\end{equation}

\section{\label{sec:mass_renormalization}Mass renormalization}
The bare lattice mass parameter \( m_{\mathrm{lat}} \) does not directly correspond to the physical fermion mass in the continuum, \( m \), due to the presence of an additive mass renormalization~\cite{Rothe2012}. The renormalized fermion mass is thus given by
\begin{equation}
\label{eq:mass_renorm}
    m_r = m_{\mathrm{lat}} + m_{\mathrm{shift}},
\end{equation}
where \( m_{\mathrm{shift}} \) denotes the mass shift. In general, \( m_{\mathrm{shift}} \) depends on the lattice spacing \( ag \), the physical volume \( Lg \), and the background electric field \( l_0 \).

For the staggered fermion discretization, an analytical prediction for the mass shift in the multi-flavor Schwinger model with \( N_F \) flavors was recently derived in Ref.~\cite{Dempsey2022}:
\begin{equation}
    \frac{m_{\mathrm{shift}}}{g} = \frac{N_F}{8\sqrt{x}}.
\end{equation}
While this mass shift vanishes in the continuum limit, its inclusion has been shown to significantly improve the convergence rate of continuum extrapolations~\cite{Dempsey2022, Angelides2023}.

For Wilson fermions, no closed-form analytical expression for the mass shift is known. A numerical method to determine this quantity within the Hamiltonian formalism was proposed in Ref.~\cite{Angelides2023}, and applied to a detailed study of the one-flavor Schwinger model. The central idea of this approach is that the vacuum expectation value of the electric field density vanishes when the renormalized fermion mass satisfies \( m_{\mathrm{r}} = 0 \). 

To determine the corresponding bare lattice mass, one computes the electric field density \( F/g \) for several small, negative values of \( m_{\mathrm{lat}}/g \) and identifies the value \( m_{\mathrm{lat}}^* \) for which the electric field vanishes. The mass shift is then defined as
\begin{equation}
    m_{\mathrm{shift}} = - m_{\mathrm{lat}}^*.
\end{equation}

This approach requires simulations at negative values of the lattice mass, which are typically inaccessible to standard Monte Carlo methods due to the sign problem. In contrast, tensor network techniques do not suffer from this limitation, allowing us to extend the method to the more challenging two-flavor Schwinger model. As in the one-flavor case, the electric field density vanishes when the renormalized mass satisfies \( m_{\mathrm{r}} = 0 \) in the two-flavor theory~\cite{Coleman1976}, enabling a direct application of the same strategy.

In the absence of a background electric field (\( l_0 = 0 \)), however, the electric field density remains zero for all values of \( m_{\mathrm{lat}} \), rendering the method inapplicable. To overcome this, one can introduce a small nonzero background field and determine the mass shift as a function of \( l_0 \), followed by an extrapolation to \( l_0 \to 0 \).

\section{\label{sec:twisted_mass}Twisted Mass fermions}
The twisted mass approach modifies the fermion mass term by introducing a chirally rotated component~\cite{Aoki1984}. When the untwisted fermion mass is tuned to zero, this setup allows for automatic \( \mathcal{O}(a) \) improvement of many physical observables, reducing leading-order lattice discretization errors to \( \mathcal{O}(a^2) \)~\cite{Frezzotti2001, Frezzotti2004}.

In this section, we first introduce the twisted mass formulation in the continuum and outline its relation to the standard (untwisted) theory. We then describe its implementation using Wilson fermions on the lattice and discuss the required tuning of the bare mass. Finally, we explain the mechanism behind automatic \( \mathcal{O}(a) \) improvement. Our discussion follows Ref.~\cite{Shindler2008}, which provides a comprehensive review of twisted mass Wilson lattice QCD. Note that the continuum discussion is framed within the Lagrangian formalism, while the lattice formulation and all numerical simulations are carried out in the Hamiltonian formalism.

\subsection{\label{sec:twisted_mass_continuum}Twisted Mass fermions in the continuum}
The fermionic part of the twisted mass Lagrangian density is given by
\begin{equation}
    \mathcal{L}_F^{\mathrm{TM}} = \sum_{f,f'=0}^{1} \bar{\chi}_f \left( i \gamma^{\mu} D_{\mu} \delta_{ff'} - (m\delta_{ff'} + i\mu\gamma^5 \tau^3_{ff'}) \right) \chi_{f'},
\end{equation}
where \(\chi_f\) is a two-component Dirac spinor with flavor index \(f\). The parameter \(\mu\) is the twisted mass, and all other symbols are as in the untwisted case. 

For a more compact expression, we introduce flavor doublets:
\begin{equation}
    \chi = \begin{pmatrix} \chi_0 \\ \chi_1 \end{pmatrix}, \quad 
    \bar{\chi} = \begin{pmatrix} \bar{\chi}_0 & \bar{\chi}_1 \end{pmatrix}.
\end{equation}
In terms of these doublets, the Lagrangian becomes:
\begin{equation}
    \mathcal{L}_F^{\mathrm{TM}} = \bar{\chi} \left( i\gamma^{\mu} D_{\mu} - (m + i\mu \gamma^5 \tau^3) \right) \chi.
\end{equation}
Here, \(\tau^3\) is the third Pauli matrix acting in flavor space, and \(\{\chi, \bar{\chi}\}\) define the so-called twisted basis.

To relate the twisted and standard formulations, the mass term can be rewritten as
\begin{equation}
    m + i\mu \gamma^5 \tau^3 = M (\cos\alpha + i\sin\alpha \gamma^5 \tau^3) = M e^{i\alpha \gamma^5 \tau^3},
\end{equation}
where \(M = \sqrt{m^2 + \mu^2}\) is the polar mass and \(\tan\alpha = \mu/m\) defines the twist angle.

Applying an axial rotation,
\begin{equation}
    \psi = e^{i\omega \gamma^5 \tau^3 / 2} \chi, \quad  
    \bar{\psi} = \bar{\chi} e^{i\omega \gamma^5 \tau^3 / 2},
\end{equation}
transforms the mass term into
\begin{equation}
    M e^{i(\alpha - \omega) \gamma^5 \tau^3},
\end{equation}
while leaving the kinetic term invariant.

Choosing \(\omega = \alpha\) rotates the theory into the so-called physical basis \(\{\psi, \bar{\psi}\}\), where the Lagrangian assumes the standard form:
\begin{equation}
    \mathcal{L}_F^{\mathrm{TM}} = \bar{\psi} (i \gamma^{\mu} D_{\mu} - M) \psi.
\end{equation}
Thus, in the continuum, the twisted mass formulation is physically equivalent to the standard formulation—just written in a rotated basis.

This equivalence also holds for lattice discretizations that preserve chiral symmetry~\cite{Frezzotti2001, Frezzotti2004}.

\subsection{\label{sec:twisted_mass_lattice}Wilson twisted mass fermions}
Wilson fermions explicitly break chiral symmetry. In this context, the twist angle \( \alpha \) parameterizes a family of lattice discretizations that share the same continuum limit. However, the specific choice of \( \alpha \) can have a beneficial impact on lattice discretization errors at finite lattice spacing. The dimensionless Hamiltonian for the two-flavor twisted mass Schwinger model is given by:
\begin{equation}
\begin{split}
    W_W^{\mathrm{TM}} &= 
    2 \sum_{n=0}^{N-1} \sum_{f=0}^{1} \left( \frac{m}{g} \sqrt{x} + x \right) (\chi_{n,f,1}^{\dag} \chi_{n,f,2} + \text{h.c.}) \\
    &\quad - 2 \frac{\mu}{g} \sqrt{x} \sum_{n=0}^{N-1} \sum_{f=0}^{1} (-1)^f (\chi_{n,f,1}^{\dag} \chi_{n,f,1} - \chi_{n,f,2}^{\dag} \chi_{n,f,2}) \\
    &\quad + 2x \sum_{n=0}^{N-2} \sum_{f=0}^{1} (\chi_{n,f,1}^{\dag} \chi_{n+1,f,2} + \text{h.c.}) \\
    &\quad + \sum_{n=0}^{N-2} \left( l_0 + \sum_{k=0}^{n} Q_k \right)^2.
\end{split}
\end{equation}

Up to the twisted mass term, this Hamiltonian is equivalent to the standard Wilson formulation. However, it is important to emphasize that \( W_W^{\mathrm{TM}} \) is written in the twisted basis, which alters the structure of certain operators—such as the first two isospin components—within this framework. The electric field density operator, however, retains the same form as in the untwisted Wilson case.

To effectively reduce lattice discretization errors, it is advantageous to work at or near maximal twist, which corresponds to a twist angle \( \alpha = \pi/2 \) in the continuum. On the lattice, this condition is realized by tuning the renormalized mass \( m_r \) to zero, or equivalently, by setting the bare lattice mass to \( m_{\mathrm{lat}} = -m_{\mathrm{shift}} \). 
Crucially, this tuning only needs to be accurate up to corrections of order \( \mathcal{O}(a) \), as required for automatic \( \mathcal{O}(a) \) improvement.

In other words, the residual renormalized mass after tuning may still be as large as \( \mathcal{O}(a) \), yet parity-even observables will exhibit leading discretization effects only at \( \mathcal{O}(a^2) \). This makes the twisted mass formulation particularly attractive in practice, as it permits improved scaling behavior without the need for extremely precise tuning. It is noteworthy that there exists an improved definition of the twist angle, ensuring robust \( \mathcal{O}(a) \) improvement without any restriction on the quark masses in the context of lattice QCD~\cite{Aoki2004}.

\subsection{\label{sec:order_a_improvement}Automatic O(a) improvement}
Automatic \(\mathcal{O}(a)\) improvement \cite{Frezzotti2001, Frezzotti2004, Aoki2004} refers to the suppression of linear lattice discretization errors such that the leading artifacts scale as \(\mathcal{O}(a^2)\), provided the fermion mass is tuned to zero, as discussed in the previous section. 

In the action (Lagrangian) formalism, this improvement can be proven via the Symanzik effective theory by analyzing the symmetries of the continuum action \cite{Aoki2006}. Specifically, it can be shown that expectation values of chirally even observables are automatically \(\mathcal{O}(a)\) improved, while chirally odd observables vanish in the continuum limit.

In the Hamiltonian formalism, a corresponding proof of \(\mathcal{O}(a)\) improvement is less straightforward due to the absence of full Euclidean spacetime symmetries. Nevertheless, we can explicitly demonstrate \(\mathcal{O}(a)\) improvement for the spectrum of the free twisted mass Dirac Hamiltonian, i.e., in the absence of gauge interactions.

For the eigenvalues of the free Hamiltonian in a finite volume \(L\) with periodic boundary conditions, we compute in App.~\ref{sec:free_fermion_spectrum}:
\begin{equation}
\begin{split}
    \lambda_k &= \pm \sqrt{ \mu^2 + m^2 + \frac{2m}{a}(1 - \cos ak) + \frac{2}{a^2}(1 - \cos ak) }, \\
    k &= \frac{2\pi n}{L}, \quad n = 0, 1, \dots, N-1.
\end{split}
\end{equation}

Expanding this expression in powers of the lattice spacing \( a \), we obtain:
\begin{equation}
|\lambda_k| = C + \frac{m k^2}{2C}\, a - \frac{3m^2 k^4 + C^2 k^4}{12 C^3}\, a^2 + \mathcal{O}(a^3),
\end{equation}
where \( C = \sqrt{\mu^2 + m^2 + k^2} \).

This expansion clearly shows that the \(\mathcal{O}(a)\) term vanishes at maximal twist (\( m = 0 \)), leaving only \(\mathcal{O}(a^2)\) corrections. Furthermore, a mistuning of the mass by an amount \(\mathcal{O}(a)\) results only in a \(\mathcal{O}(a^2)\) deviation in the spectrum, confirming the robustness of the improvement mechanism.

\section{\label{sec:mps}Matrix product states}
Matrix Product States (MPS) are an entanglement-based ansatz for describing one-dimensional quantum many-body systems~\cite{Schollwoeck_2011}. For a system with \(N\) spins \(\sigma_i\) on a lattice with open boundary conditions, the MPS representation of the system's state \(|\psi\rangle\) is given by:  
\begin{equation}
    |\psi\rangle = \sum_{\sigma_0,\dots,\sigma_{N-1}} A_0^{\sigma_0} A_1^{\sigma_1} \cdots A_{N-1}^{\sigma_{N-1}} 
    |\sigma_0\rangle \otimes |\sigma_1\rangle \otimes \cdots \otimes |\sigma_{N-1}\rangle.
\end{equation}  

For \(0 < k < N-1\), each tensor \( A_k^{\sigma_k} \) is a complex \( D \times D \) matrix associated with a specific spin configuration \(\sigma_k\). The parameter \( D \), known as the bond dimension, governs the number of variational parameters and determines the maximum amount of entanglement the MPS can faithfully represent. The boundary tensors \( A_0^{\sigma_0} \) and \( A_{N-1}^{\sigma_{N-1}} \) are row and column vectors of dimension \( D \), respectively, ensuring that the full tensor contraction yields a complex scalar coefficient for each basis state.

The bond dimension \( D \) is directly related to the entanglement structure of the state. To quantify this, consider a bipartition of the system into two contiguous subsystems, \( A \) and \( B \), and define the entanglement entropy as the von Neumann entropy of the reduced density matrix \(\rho_A = \mathrm{Tr}_B |\psi\rangle\langle\psi|\):  
\begin{equation}
    S_A = -\mathrm{Tr}(\rho_A \log \rho_A).
\end{equation}

For an MPS with bond dimension \( D \), the maximum entanglement entropy across any bipartition is bounded by:
\begin{equation}
    S_A \leq \log D.
\end{equation}
This means that MPS with small bond dimensions are well suited for representing weakly entangled states, such as ground states of gapped one-dimensional local Hamiltonians, which obey an area law for entanglement. In contrast, highly entangled states, such as those in critical systems, may require significantly larger bond dimensions for accurate representation.

To approximate ground states of the Hamiltonians considered in our study, we employ the two-site variational ground state search implemented in the \texttt{ITensors} library \cite{Fishman2022}. This algorithm iteratively updates pairs of tensors \( A_i^{\sigma_i} \) while keeping the remaining tensors fixed, minimizing the energy expectation value \( \langle \psi | W | \psi \rangle \). A full update from one boundary to the other and back is referred to as a \emph{sweep}. The algorithm is terminated once the relative change in the energy expectation value falls below a threshold of \( \eta = 10^{-10} \), which corresponds to a relative uncertainty of approximately \( \sqrt{\eta} \) for the electric field density~\cite{Haegeman2011}.

Although tensor network algorithm can directly deal with fermionic degrees of freedom~\cite{Pineda2010,Kraus2010,Corboz2010}, in $(1+1)$-dimensions it is convenient to translate them to spins by applying a Jordan-Wigner transformation. This transformation simplifies the implementation and improves computational efficiency by allowing us to work entirely within the well-established spin framework of tensor network libraries.

The same algorithm can also be used to estimate excited states by minimizing the expectation value of a modified effective Hamiltonian,
\begin{equation}
    W_{\mathrm{eff}} = W + \lambda \sum_i |E_i| |\psi_i\rangle \langle \psi_i|,
\end{equation}
where \( |\psi_i\rangle \) are the previously computed low-lying states and \( E_i \) their corresponding energies. The tunable parameter \( \lambda \) controls the strength of the penalty for overlap with lower-lying states, thereby suppressing ground state components in the variational search.

An additional important aspect of our implementation is the restriction of the MPS to the sector of vanishing total charge, $\sum_n Q_n = 0$. This is achieved by explicitly incorporating the symmetry into the MPS ansatz.

The bond dimension of the MPS was chosen in the range \( D = 100 \) to \( D = 2000 \), depending on the excitation level, fermion discretization, and system size. For the smallest bond dimension, the initial MPS was initialized randomly. For larger bond dimensions, we used the optimized states from previous runs with lower bond dimension as initial states. This warm-start strategy significantly reduces the computational cost, as simulations at lower bond dimension are much faster while still providing a good approximation to the ground state.

\begin{figure}
    \centering
    \includegraphics[width=0.48\textwidth]{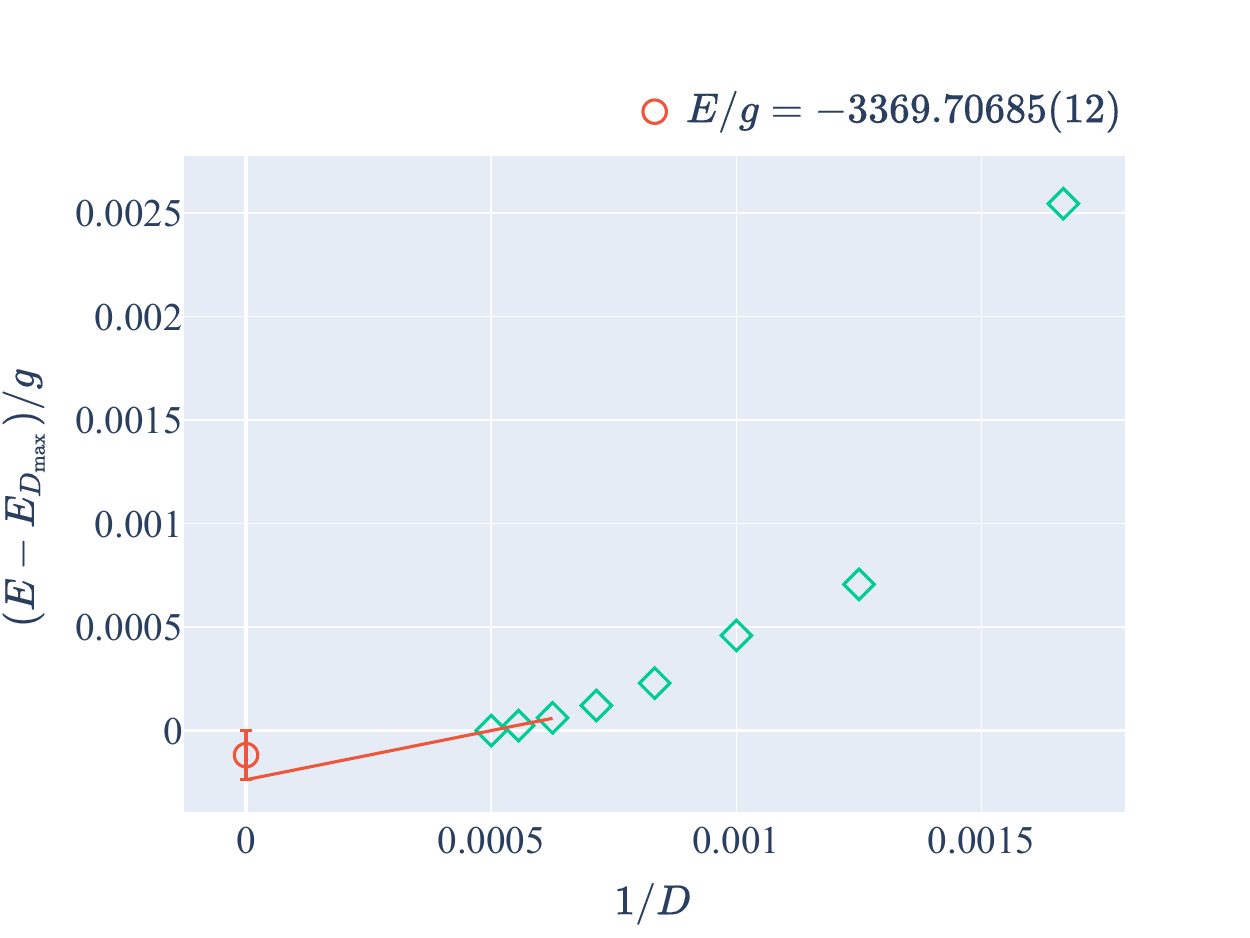}
    \caption{Convergence of the ground state energy for twisted mass fermions at maximal twist as a function of bond dimension \(D\), for \(N = 200\), \(L g = 30\), and \(\mu/g = 0.1\). Shown is the difference between the computed energy at a given bond dimension and the value at the maximum bond dimension \(D_{\text{max}}\). The final extrapolated value is indicated by a red circle, with its numerical value shown in the legend.}
    \label{fig:extrapolation_D}
\end{figure}

To estimate the uncertainty of the numerical method, we implement the following strategy. We consider the values of observables obtained for bond dimensions larger than 250, as these are typically already close to their converged values. If a strictly monotonic behavior in \( 1/D \) is observed, we perform a linear extrapolation using the three largest bond dimension values in \( 1/D \). The final value is then taken as the mean of the extrapolated value and the result at the largest bond dimension, with an associated uncertainty given by half the difference between the two. An illustration of this procedure is shown in Fig.~\ref{fig:extrapolation_D}.

If, on the other hand, a non-monotonic dependence on the bond dimension is observed, we take the result at the largest bond dimension as the final value and assign an uncertainty that encompasses the full spread of values within the non-monotonic region. The total uncertainty is obtained by combining this finite bond dimension uncertainty with the uncertainty from the energy convergence stopping criterion in quadrature.

\section{\label{sec:results}Numerical results}

We begin by analyzing the scaling behavior of the electric field density to demonstrate automatic \( \mathcal{O}(a) \) improvement for twisted mass fermions. Next, we determine the mass renormalization and study its impact on the scaling behavior. We then investigate the low-energy spectrum for all three fermion discretizations, with particular focus on the pion mass, which we extract using both dispersion relation fits and finite-volume scaling. All simulations are performed in the presence of a large background electric field, corresponding to \( l_0 = \theta/(2\pi) = 0.4 \), which introduces a sign problem for Monte Carlo approaches.

\subsection{\label{sec:electric_field_density}Electric field density}

The electric field density is a fundamental observable in the Schwinger model, directly reflecting the gauge dynamics and vacuum structure of the theory. As a local, gauge-invariant quantity, it is sensitive to confinement and vacuum polarization effects, and serves as a stringent test for the accuracy of numerical methods and discretizations. In particular, its scaling behavior with lattice spacing provides insight into discretization errors and the effectiveness of improvement schemes. Despite its physical relevance, the electric field density lacks exact analytical predictions in the massive two-flavor case, making it an ideal target for precision tensor network studies.

As a first step, we investigate the scaling behavior of the electric field density at fixed lattice fermion masses \( m_{\mathrm{lat}}/g \). Previous studies of the Schwinger model using tensor network methods have successfully performed continuum extrapolations without applying mass renormalization. This approach avoids the computational overhead associated with determining mass counterterms and eliminates the risk of introducing related systematic errors, thereby enabling a clean and precise study of scaling behavior.

To assess how different fermion discretizations approach the continuum limit, we consider a fixed physical volume of \( Lg = N/\sqrt{x} = 30 \) and simulate system sizes ranging from 130 to 200 sites in steps of ten. Accounting for the number of fermion flavors, this corresponds to systems with up to 400 spins in the staggered formulation and up to 800 spins for Wilson and twisted mass fermions. For the fixed volume, these system sizes correspond to lattice spacings as small as \( ag = 0.15 \).

In Fig.~\ref{fig:combined_fits_efd}, we present our numerical results for lattice fermion masses \( m_{\mathrm{lat}}/g = 0.1, 0.2, 0.4 \). The small error bars of the individual data points reflect the uncertainty due to finite bond dimension effects, as described in Sec.~\ref{sec:mps}. To investigate the continuum limit in the absence of exact analytical predictions, we employ the following strategy:

\begin{figure*}[ht] 
    \centering

    \includegraphics[width=1.0\textwidth]{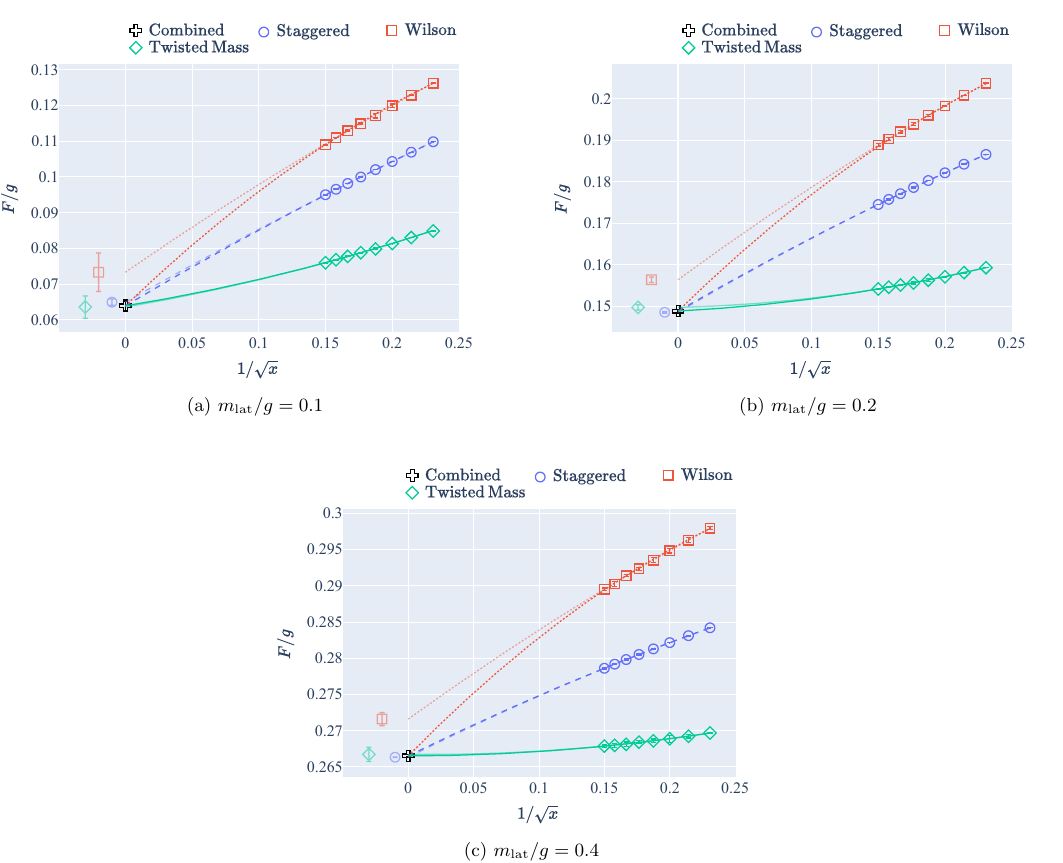}
    \caption{Scaling behavior of the electric field density \( F/g \) as a function of the lattice spacing \( 1/\sqrt{x} \). Faint lines show individual quadratic fits for each fermion discretization, while bold lines represent the combined fit with a shared continuum limit. The continuum extrapolation points from the individual fits are horizontally offset from zero for better visibility.}
    \label{fig:combined_fits_efd}
\end{figure*}

First, we fit the data points for each fermion discretization separately using a quadratic function of the form \( A (1/\sqrt{x})^2 + B/\sqrt{x} + C \) and extrapolate to vanishing lattice spacing \( 1/\sqrt{x} = 0 \). The individual fits are shown as faint lines in Fig.~\ref{fig:combined_fits_efd}. For improved visibility, the extrapolated continuum values are slightly offset from zero.

To estimate the uncertainty of the extrapolation, we employ a jackknife resampling procedure~\cite{Efron1982}. Let \( x^{(i)} \) denote the extrapolated continuum value obtained by leaving out the \( i \)-th data point from the fit, and let \( M \) be the number of such jackknife samples (equal to the number of data points). The jackknife estimate of the mean is
\begin{equation}
\bar{x}_{\text{jack}} = \frac{1}{M} \sum_{i=0}^{M-1} x^{(i)},
\end{equation}
and the corresponding variance is given by
\begin{equation}
\sigma^2_{\text{jack}} = \frac{M - 1}{M} \sum_{i=0}^{M-1} \left( x^{(i)} - \bar{x}_{\text{jack}} \right)^2.
\end{equation}

We report the full-sample fit result as the final extrapolated value, and assign the jackknife standard deviation as its uncertainty.

Next, we perform a combined quadratic fit across all three fermion discretizations, constrained to a shared continuum value \( C \), shown by the bold lines in Fig.~\ref{fig:combined_fits_efd}. This approach yields a stable and precise continuum estimate that serves as a reference point against which the individual fits can be compared. The uncertainty is again estimated using the jackknife method, this time with each stripped dataset omitting three points—one for each fermion discretization. For all three masses, the resulting uncertainties are smaller than the marker size in the figure.

Across all masses, twisted mass fermions exhibit the fastest convergence to the continuum, followed by staggered fermions. Wilson fermions show the slowest convergence. The individual fits for staggered and twisted mass fermions are consistent with the combined fit within uncertainties, whereas the individual fit for Wilson fermions slightly overestimates the continuum value compared to both the other individual extrapolations and the combined extrapolation.

More specifically, the Wilson fits appear almost linear over the accessible range of lattice spacings, while the combined fit—which provides a more reliable estimate of the continuum limit—displays a non-negligible curvature. This discrepancy indicates that a broader range of lattice spacings, and in particular finer lattice spacings, is necessary to control discretization effects when using Wilson fermions. These observations highlight the practical advantage of improved fermion discretizations, such as twisted mass fermions, which enable more reliable continuum extrapolations at comparable lattice spacings and computational cost.

To support these qualitative observations, we present the fit parameters from the combined continuum extrapolation in Tab.~\ref{tab:combined_fits_efd}. The quoted uncertainties are 95\% confidence intervals obtained via the jackknife procedure.

\begin{table*}[ht]
  \centering
  \renewcommand{\arraystretch}{1.2} 
  \setlength{\tabcolsep}{10pt}       
  \begin{tabular}{llcccccc}
    \toprule
    Mass & \begin{tabular}[c]{@{}l@{}}Fermion\\ Type\end{tabular} & A (\((1/\sqrt{x})^2\)) & B (\(1/\sqrt{x}\)) & C (1) & \begin{tabular}[c]{@{}c@{}}Linear\\ Contribution\end{tabular} & \begin{tabular}[c]{@{}c@{}}Quadratic\\ Contribution\end{tabular} \\
    \midrule
    \multirow{3}{*}{0.1} 
        & Staggered    & \(-0.099_{-0.024}^{+0.017}\)   & \(0.22_{-0.01}^{+0.01}\) & & \(0.92_{-0.02}^{+0.01}\) & \(0.08_{-0.01}^{+0.02}\) \\
        & Wilson    & \(-0.383_{-0.027}^{+0.021}\)   & \(0.36_{-0.01}^{+0.01}\)   & \(0.0639_{-0.0009}^{+0.0007}\) & \(0.835_{-0.004}^{+0.002}\) & \(0.165_{-0.002}^{+0.004}\) \\
        & Twisted Mass    & \(0.134_{-0.028}^{+0.015}\)   & \(0.06_{-0.01}^{+0.02}\)  & & \(0.71_{-0.06}^{+0.08}\) & \(0.29_{-0.08}^{+0.06}\) \\
    \midrule
    \multirow{3}{*}{0.2} 
        & Staggered    & \(-0.096_{-0.003}^{+0.009}\)   & \(0.186_{-0.004}^{+0.001}\)   &  & \(0.913_{-0.002}^{+0.004}\) & \(0.087_{-0.004}^{+0.002}\) \\
        & Wilson    & \(-0.334_{-0.012}^{+0.019}\)   & \(0.315_{-0.006}^{+0.003}\)   & \(0.1488_{-0.0002}^{+0.0004}\) & \(0.836_{-0.003}^{+0.004}\) & \(0.164_{-0.004}^{+0.003}\) \\
        & Twisted Mass    & \(0.12_{-0.01}^{+0.02}\)   & \(0.017_{-0.004}^{+0.002}\)   & & \(0.43_{-0.09}^{+0.05}\) & \(0.57_{-0.05}^{+0.09}\) \\
    \midrule
    \multirow{3}{*}{0.4} 
        & Staggered    & \(-0.0505_{-0.0028}^{+0.0015}\)   & \(0.088_{-0.001}^{+0.002}\)   & & \(0.904_{-0.006}^{+0.003}\) & \(0.096_{-0.003}^{+0.006}\) \\
        & Wilson    & \(-0.21_{-0.02}^{+0.01}\)   & \(0.184_{-0.001}^{+0.004}\)   & \(0.2665_{-0.0002}^{+0.0001}\) & \(0.827_{-0.006}^{+0.004}\) & \(0.173_{-0.004}^{+0.006}\) \\
        & Twisted Mass    & \(0.0581_{-0.0022}^{+0.0013}\)   & \(0.000_{-0.001}^{+0.002}\)   &  & \(0.02_{-0.02}^{+0.09}\) & \(0.98_{-0.09}^{+0.02}\) \\
    \bottomrule
  \end{tabular}
  \caption{Constrained fits for the electric field density: Comparison of coefficients for \((1/\sqrt{x})^2\) (A), \(1/\sqrt{x}\) (B), and the constant (C), along with the corresponding linear and quadratic contributions for masses 0.1, 0.2, and 0.4 for different fermion types. The fits are given in the form \(A^{+\text{err}_{\text{up}}}_{-\text{err}_{\text{low}}}(1/\sqrt{x})^2 + B^{+\text{err}_{\text{up}}}_{-\text{err}_{\text{low}}}(1/\sqrt{x}) + C^{+\text{err}_{\text{up}}}_{-\text{err}_{\text{low}}}\).}
  \label{tab:combined_fits_efd}
\end{table*}

While the fit coefficients \(A\), \(B\), and \(C\) characterize the extrapolation function, their numerical values alone do not convey the relative importance of each term over the finite range of lattice spacings considered. For instance, a larger quadratic coefficient \(A\) can have less impact than a smaller linear coefficient \(B\) if the corresponding powers of \(1/\sqrt{x}\) differ significantly across the data range.

To quantify this, we define the relative contributions of the linear and quadratic terms within the sampled data range as
\begin{equation}
\begin{split}
    \text{Linear Contribution:} \quad & 
    \frac{|B \langle 1/\sqrt{x} \rangle|}{|B \langle 1/\sqrt{x} \rangle| + |A \langle (1/\sqrt{x})^2 \rangle|}, \\
    \text{Quadratic Contribution:} \quad & 
    \frac{|A \langle (1/\sqrt{x})^2 \rangle|}{|B \langle 1/\sqrt{x} \rangle| + |A \langle (1/\sqrt{x})^2 \rangle|},
\end{split}
\end{equation}
where the averages \(\langle \cdot \rangle\) are taken over the values of \(1/\sqrt{x}\) used in the fit. These ratios offer a more intuitive measure of the relative influence of each term across the range of data points included in the extrapolation. Uncertainties are estimated using the jackknife method.

The main observation is that staggered and Wilson fermions exhibit predominantly linear discretization errors across all fermion masses considered. For staggered fermions, the linear contribution exceeds 0.9 throughout the data range, while for Wilson fermions it remains slightly lower, at approximately 0.83. This behavior appears largely insensitive to variations in the fermion mass within the explored parameter regime. In contrast, twisted mass fermions display a distinct scaling pattern: the quadratic contribution becomes increasingly dominant as the fermion mass increases, rising from 0.29 at \( m_{\mathrm{lat}}/g = 0.1 \) to 0.98 at \( m_{\mathrm{lat}}/g = 0.4 \). For the largest mass, the linear coefficient is zero within its confidence interval, indicating that the data is consistent with a purely quadratic scaling in this regime.

This behavior can be interpreted in the context of the automatic \( \mathcal{O}(a) \) improvement inherent in the twisted mass formulation. When mass renormalization is omitted and the untwisted lattice fermion mass is set to zero, increasing the twisted mass parameter effectively tunes the system toward maximal twist. At maximal twist, theoretical arguments guarantee the cancellation of leading-order discretization effects (\( \mathcal{O}(a) \)) for parity-even, multiplicatively renormalizable observables, based on the symmetry structure of the Symanzik effective theory in the Lagrangian formalism.

A corresponding analysis in the Hamiltonian framework is considerably more subtle. In particular, the lack of full Euclidean space-time symmetry complicates the derivation of an analogous Symanzik expansion. For this reason, we do not attempt a formal proof of automatic \( \mathcal{O}(a) \) improvement here. Instead, we rely on explicit numerical evidence to support the presence of this improvement mechanism in the Hamiltonian setting. 

While the electric field density is formally a gauge observable and thus not directly covered by standard fermionic improvement arguments, it is a non-dynamical quantity in the Schwinger model, entirely determined by the distribution of fermionic charges. In particular, it can be expressed as a sum over local charge densities. From this perspective, the electric field density inherits its discretization behavior from the underlying fermionic dynamics. Therefore, even though it lies outside the class of observables traditionally addressed by standard improvement proofs, our numerical results suggest that it exhibits effective \( \mathcal{O}(a^2) \) scaling in practice when computed with twisted mass fermions. This indicates that the improvement mechanism may extend to a broader class of observables in specific models, such as QED in 1+1 dimensions, and warrants further theoretical investigation.

\subsection{\label{sec:mass_renormalization_results}Mass renormalization}
In Sec.~\ref{sec:mass_renormalization}, we highlighted the usefulness of mass renormalization for improving the convergence to the continuum limit. In the case of twisted mass fermions, subtracting the additive mass shift to set the untwisted mass to zero corresponds to tuning the theory to maximal twist. Within the Lagrangian formalism, this is typically achieved by enforcing a vanishing PCAC mass, defined via an axial Ward identity. However, since the definition of the PCAC mass relies on a symmetry relation that does not hold in the Hamiltonian formulation, we instead adopt the alternative procedure proposed in Ref.~\cite{Angelides2023} and summarized in Sec.~\ref{sec:mass_renormalization}, originally introduced in the context of Wilson fermions. The method is based on identifying the value of the lattice mass \( m_{\mathrm{lat}}^* \) for which the electric field density vanishes.

To accurately determine this zero crossing, we compute the electric field density for several values of \( m_{\mathrm{lat}} \) in the vicinity of the expected crossing point and fit a quadratic function to the data. The mass shift \( m_{\mathrm{shift}} \) is then defined as the root of this fitted function. Its uncertainty is estimated using the jackknife method, following the same procedure employed for estimating the uncertainty of the continuum extrapolation in the previous section.

The mass shift exhibits a strong dependence on the lattice spacing. Consequently, we compute it separately for each lattice spacing considered in this study, for both staggered and Wilson fermions. In Fig.~\ref{fig:mass_shift_N200}, we illustrate the previously described procedure for the case of Wilson fermions at the finest lattice spacing \( 1/\sqrt{x} = 0.15 \). In addition to the electric field density, we also monitor the entanglement entropy. This leads to the following observations:

\begin{figure}
    \centering
    \includegraphics[width=0.48\textwidth]{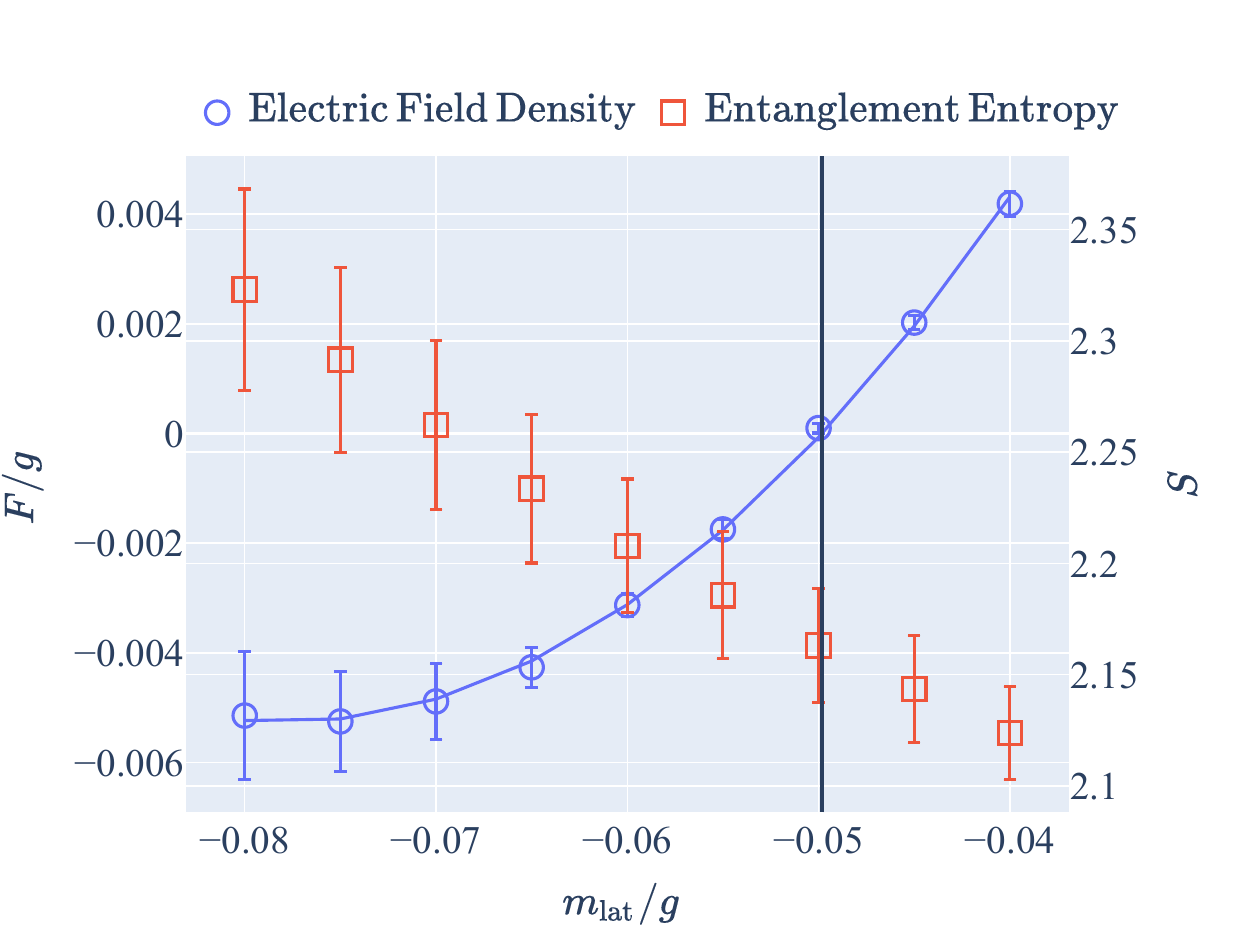}
    \caption{Electric field density \( F/g \) and entanglement entropy \( S \) as functions of the lattice mass \( m_{\mathrm{lat}}/g \) for Wilson fermions at the smallest lattice spacing \( 1/\sqrt{x} = 0.15 \). The first zero crossing of \( F/g \), indicated by the black vertical line, can be used to determine the additive mass renormalization on the lattice.}
    \label{fig:mass_shift_N200}
\end{figure}

The quadratic fit provides an accurate description of the electric field density across the scanned mass range and allows for a reliable determination of the zero crossing point. As the lattice fermion mass becomes more negative, the entanglement entropy increases. As discussed in Sec.~\ref{sec:mps}, higher entanglement entropy requires a larger bond dimension in the MPS ansatz to maintain numerical accuracy, thereby increasing computational complexity. This is reflected in the larger uncertainties observed in the electric field density in the regime of high entanglement, where the bond dimension becomes a limiting factor.

During the computation of the mass shift, we observed mixing between the ground state and excited states, as indicated by a non-vanishing isospin \( J^2 \). This mixing complicates the reliable extraction of the mass shift and points to a small or vanishing energy gap between the ground state and nearby excitations. Such behavior is reminiscent of the phase structure in lattice QCD with Wilson fermions, where two critical points enclose an intermediate phase characterized by a vanishing gap. Interestingly, for both Wilson and staggered fermions, we observed two distinct zero crossings of the electric field density, further suggesting the presence of a nontrivial phase structure in this regime.

To ensure robust ground state extraction and suppress undesired mixing, we enforce vanishing isospin \( J^2 = 0 \) by adding a penalty term to the Hamiltonian. This modification stabilizes the ground state and enables a reliable determination of the mass shift.

\begin{figure}
    \centering
    \includegraphics[width=0.48\textwidth]{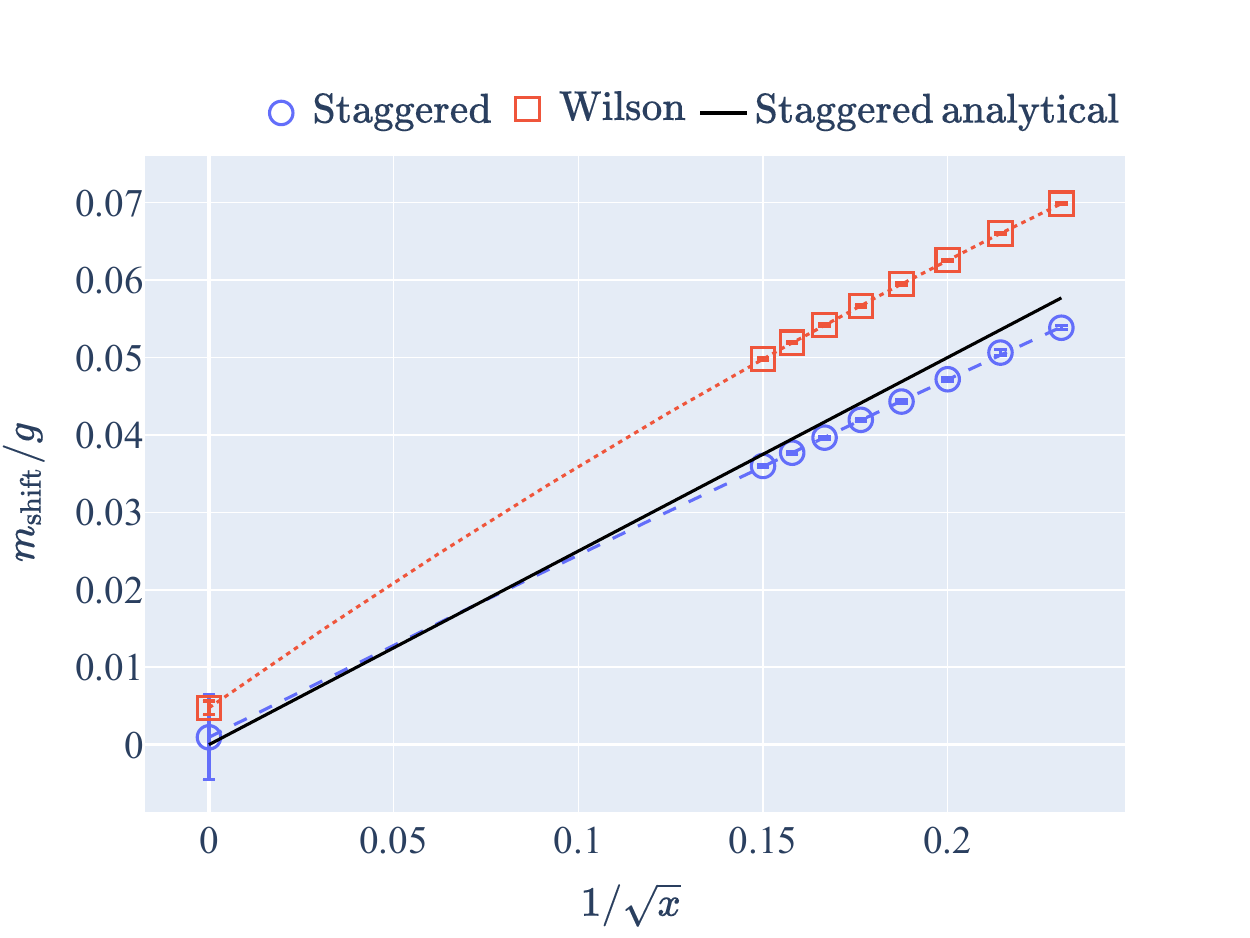}
    \caption{Additive mass renormalization \( m_{\mathrm{shift}}/g \) as a function of the lattice spacing \( 1/\sqrt{x} \), extracted from the zero crossing of the electric field density for staggered and Wilson fermions. Dashed lines indicate quadratic continuum extrapolations. The solid black line represents the analytical prediction for staggered fermions with periodic boundary conditions from Ref.~\cite{Dempsey2022}.}
    \label{fig:mass_shift_overview}
\end{figure}

In Fig.~\ref{fig:mass_shift_overview}, we present the mass shift estimates for all lattice spacings considered in our study for staggered and Wilson fermions. We also include the analytical prediction for staggered fermions from Ref.~\cite{Dempsey2022}, allowing us to provide a finite-volume, open-boundary test of this prediction in the two-flavor Schwinger model. For staggered fermions, we observe good agreement with the theoretical prediction, with the agreement improving at smaller lattice spacings. As expected, we verify that the mass shift vanishes in the continuum limit by fitting a quadratic function to the data and estimating uncertainties using the procedure described in Sec.~\ref{sec:electric_field_density}.

For Wilson fermions, the mass shift is significantly larger than for staggered fermions across all lattice spacings. The quadratic fit extrapolates to zero in the continuum limit, with a small residual deviation that can be attributed to the limited fitting range, as previously observed in the case of the electric field density for Wilson fermions. To further validate the mass shift estimate and assess its impact, we will compare continuum extrapolations of the electric field density with and without including mass renormalization.

\begin{figure*}[ht] 
    \centering

    \includegraphics[width=1.0\textwidth]{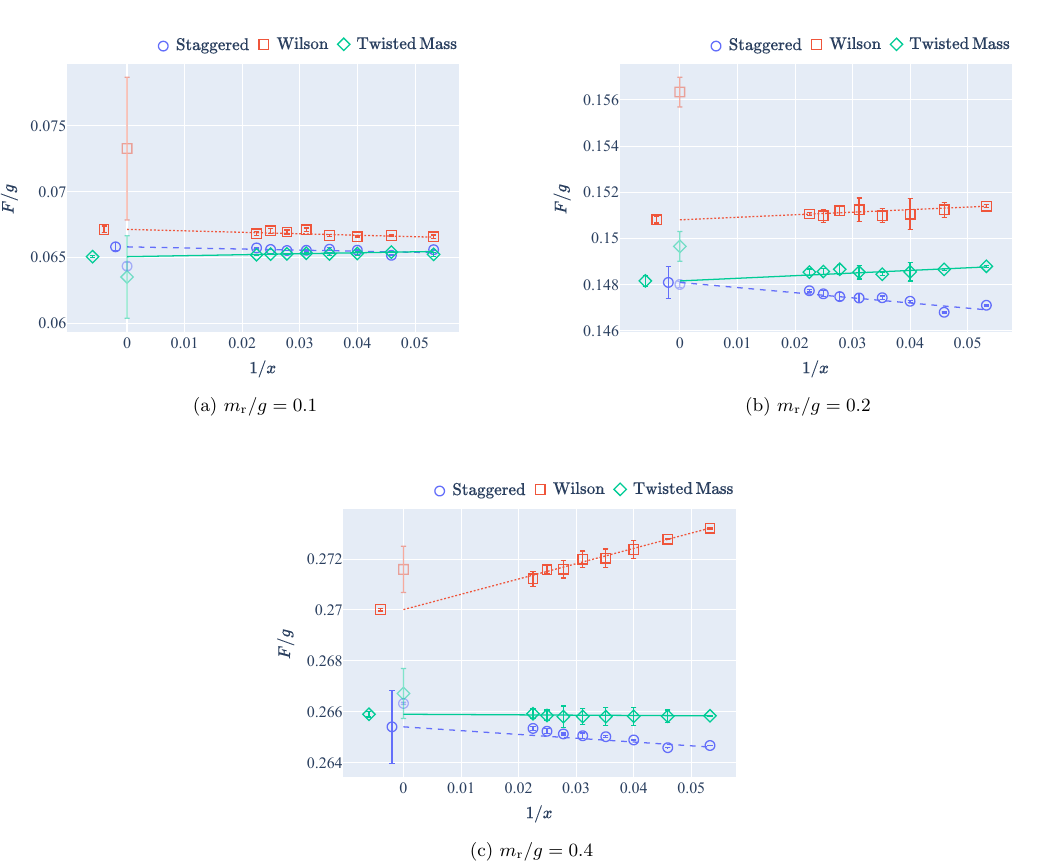}
    \caption{Scaling behavior of the electric field density \( F/g \) at fixed renormalized mass \( m_{\mathrm{r}}/g \). Bold lines indicate purely quadratic fits for all three fermion discretizations; the continuum extrapolation points are again horizontally offset for clarity. Faint markers show the previously obtained extrapolations at fixed lattice mass. Note the significantly smaller y-axis scale compared to Fig.~\ref{fig:combined_fits_efd}.}
    \label{fig:efd_quadratic_only}
\end{figure*}

In Fig.~\ref{fig:efd_quadratic_only}, we compare purely quadratic fits of the form \( A (1/\sqrt{x})^2 + C \) (bold lines) obtained at fixed renormalized mass to the previously obtained continuum extrapolations at fixed lattice mass (faint markers). As before, the continuum extrapolations are slightly offset from zero to improve readability. For all three masses considered, we observe that the inclusion of mass renormalization significantly enhances the rate at which the continuum limit is approached (note the much smaller $y$-axis scale for the electric field density). This behavior is consistent with the findings reported in Refs.~\cite{Dempsey2022, Angelides2023} for the one-flavor case.

In particular, the systematic overestimation observed for Wilson fermions at fixed lattice mass is significantly reduced after mass renormalization, bringing the results into much closer agreement with those of the other two fermion discretizations. This supports our interpretation that the deviations seen in the unrenormalized case are primarily due to the limited range of accessible lattice spacings. The improvement achieved upon renormalization further underscores the practical advantages of using improved fermion discretizations for reliable continuum extrapolation.

For the two larger masses, we find perfect agreement between staggered and twisted mass fermions within their respective uncertainties. For the smallest mass, we observe a very small relative deviation below 1\%, which disappears at larger masses. This suggests a minor systematic bias introduced by the heuristic method used to determine the mass renormalization.

Across all three masses, the continuum values obtained from Wilson fermions are consistently slightly larger than those from the other two discretizations. This deviation is very small with a maximum relative deviation of approximately 1\%. 

\subsection{\label{sec:mass_spectrum}Mass spectrum}
\begin{figure*}[ht] 
    \centering
    \includegraphics[width=1.0\textwidth]{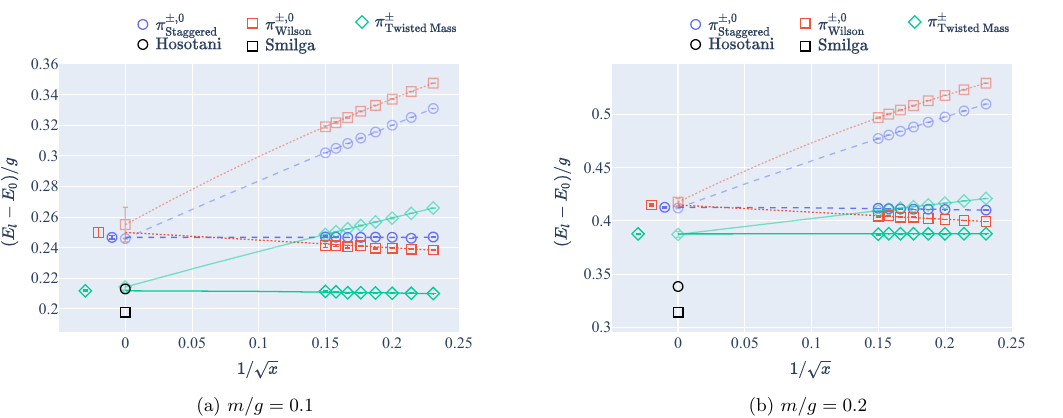}
    \caption{Scaling behavior of the energy gap \((E_l - E_0)/g\) of the pion for the three different fermion discretizations. Faint lines show quadratic fits to the data obtained at fixed lattice mass, while bold lines show purely quadratic fits for data at fixed renormalized mass. As in previous figures, some of the extrapolated values are offset from zero for clarity. Analytical estimates from Ref.~\cite{Hosotani1998} (Hosotani) and Ref.~\cite{Smilga1997} (Smilga) for the pion mass are included for comparison. Note that the energy gap does not directly correspond to the pion mass without accounting for additional systematic effects.}
    \label{fig:mass_gap_ms_vs_no_ms}
\end{figure*}

Having studied the scaling behavior of the electric field density in the ground state, we now turn to the low-lying mass spectrum of the two-flavor Schwinger model for the three fermion discretizations. We begin by analyzing the scaling behavior of the energy gap between the ground state and the first excited state. As in the case of the electric field density, we present quadratic fits for the data without mass renormalization (faint lines) and purely quadratic fits for the data including mass renormalization (bold lines). Results are shown in Fig.~\ref{fig:mass_gap_ms_vs_no_ms} for \( m/g = 0.1 \) and \( m/g = 0.2 \).

Several observations are noteworthy. The extrapolated data with and without mass renormalization agree well for each fermion discretization individually. The inclusion of the renormalization significantly improves the convergence toward the continuum limit. For both masses, the energy gap obtained from staggered and Wilson fermions matches precisely, while the energy gap from twisted mass fermions lies significantly lower. This suggests that the energy gap is strongly affected by additional systematic effects, most notably finite volume corrections, which must be carefully addressed in order to reliably extract the pion mass from the excitation gap.

For the smaller fermion mass, the energy gap obtained from twisted mass fermions matches the analytical estimate for the mass gap from Ref.~\cite{Hosotani1998} with great precision. This indicates that the continuum extrapolation using twisted mass fermions is less sensitive to finite-volume effects and other systematic uncertainties. For the larger mass, the energy gaps extracted from staggered and Wilson fermions lie above those from twisted mass fermions, and all three discretizations yield continuum values exceeding the analytical prediction. It is important to note, however, that the analytical estimates of Refs.~\cite{Smilga1997, Hosotani1998} are derived under the assumption of small fermion masses, which may explain the observed discrepancy with the numerical results at larger masses.

We now turn to analyzing the systematic effects that influence the extraction of the pion mass. To keep the computational cost manageable, we focus on a fixed lattice spacing of \( 1/\sqrt{x} = 0.273 \), corresponding to a system size of \( N = 110 \) sites and a renormalized fermion mass \( m_{\mathrm{r}}/g = 0.1 \). Within the Hamiltonian formulation of the Schwinger model, two principal approaches are commonly employed in the literature to extract the pion mass from the excitation gap.

The first approach involves taking the thermodynamic limit \( N \to \infty \) at fixed lattice spacing before performing the continuum extrapolation \cite{Banuls2013}. This is equivalent to taking the infinite-volume limit, as the physical volume is given by \( Lg = N/\sqrt{x} \). While this method provides a clean separation of infrared and ultraviolet effects, it requires simulations at multiple large volumes, making it computationally expensive.

The second approach extracts the pion mass from a fit to the dispersion relation \cite{Itou2023, Itou2024}. In this method, multiple excited states are computed to reconstruct the pion's dispersion curve. While this avoids the need for multiple volume simulations, it introduces its own complications: it relies on a pseudo-momentum operator that does not commute with the Hamiltonian when open boundary conditions are used. As a result, although this method has been successfully applied, it may be less suitable for high-precision computations due to its reliance on approximate quantum numbers.

\begin{figure*}[ht] 
    \centering

    \includegraphics[width=1.0\textwidth]{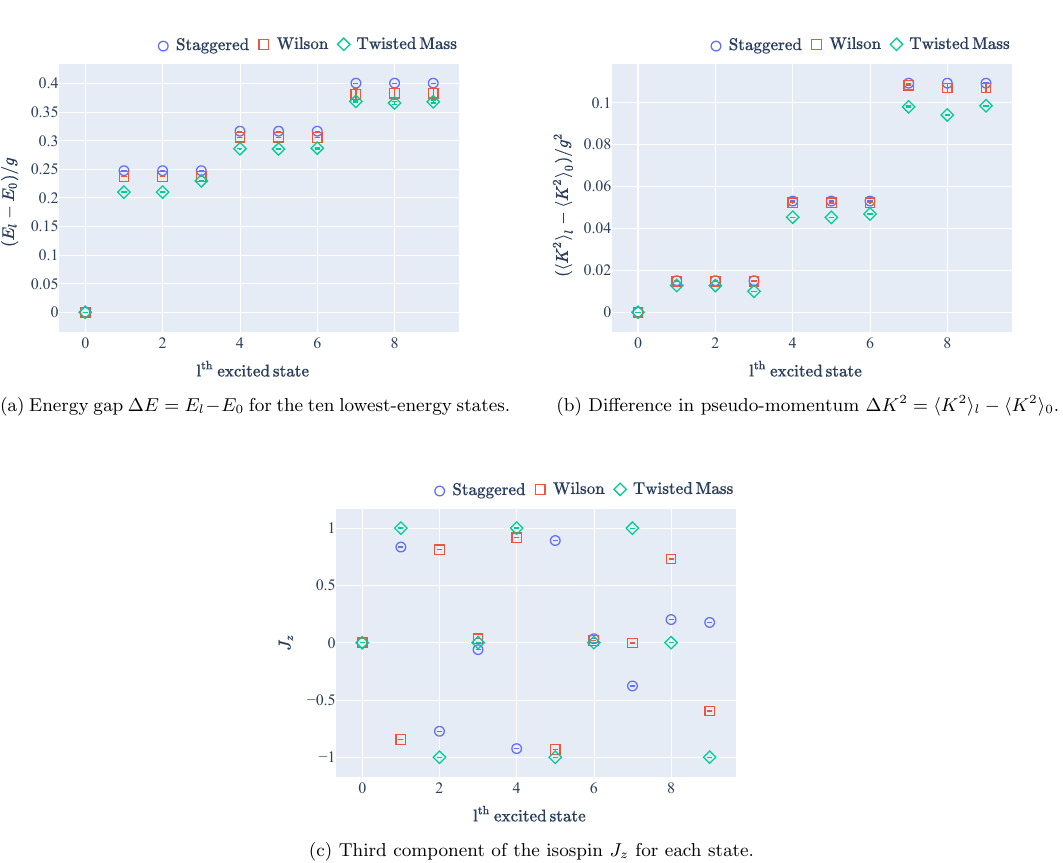}
    \caption{Key observables for the ten lowest-energy states of the massive two-flavor Schwinger model, computed using a variational ground state search with staggered, Wilson, and twisted mass fermions. Panels show (a) the energy gap relative to the ground state, (b) the pseudo-momentum gap, and (c) the third component of the isospin \( J_z \). The explicit isospin breaking in the twisted mass formulation gives rise to distinct features, which are discussed in the main text.}
    \label{fig:spectrum}
\end{figure*}

To compare these two approaches, we begin by analyzing the low-lying spectrum of the theory. In Fig.~\ref{fig:spectrum}, we show the energy gap between the ground state and the nine lowest excited states, the corresponding pseudo-momentum differences, and the third component of isospin \( J_z \). Following the ground state, we observe three triplets which correspond to momentum excitations of the pion. For staggered and Wilson fermions, these triplet states are degenerate in both energy and pseudo-momentum, reflecting the intact isospin symmetry at finite lattice spacing. In contrast, for twisted mass fermions, isospin symmetry is explicitly broken at finite lattice spacing. This breaking is most pronounced in the first triplet, where the charged pions appear at lower energy than the neutral pion, which in turn carries less pseudo-momentum. While this energy splitting is sizable—around 20\% at this lattice spacing—it is expected to vanish quadratically in the continuum limit due to automatic \( \mathcal{O}(a) \) improvement.

Isospin breaking also manifests in the structure of the variationally obtained eigenstates. In the twisted mass formulation, the ground state search yields pion states that are essentially exact eigenstates of \( J_z \). For the first two triplets, the algorithm finds the charged pions first, while for the third triplet it finds one charged pion, then the neutral pion, and finally the second charged pion. For staggered and Wilson fermions, on the other hand, the variational search yields states that are admixtures of multiple isospin components. 

\subsection{\label{sec:pion_mass_dispersion}Pion mass from the dispersion relation}

Having computed the energy gaps \( \Delta E_l = E_l - E_0 \) and the pseudo-momentum estimates \( \Delta K^2_l = \langle K^2 \rangle_l - \langle K^2 \rangle_0 \), we proceed to determine the pion masses via the dispersion relation, following the approach of Ref.~\cite{Itou2023}. At finite lattice spacing, the continuum dispersion relation is modified by lattice artifacts. To account for these effects, we perform dispersion fits using the form
\begin{equation}
    \Delta E = \sqrt{m_{\pi}^2 + b^2 \Delta K^2},
\end{equation}
where the parameter \( b \) effectively captures deviations from the continuum relation induced by discretization effects. The pion mass \( m_{\pi} \) is then extracted by extrapolating to zero pseudo-momentum, \( \Delta K^2 \to 0 \).

\begin{figure}
    \centering
    \includegraphics[width=0.48\textwidth]{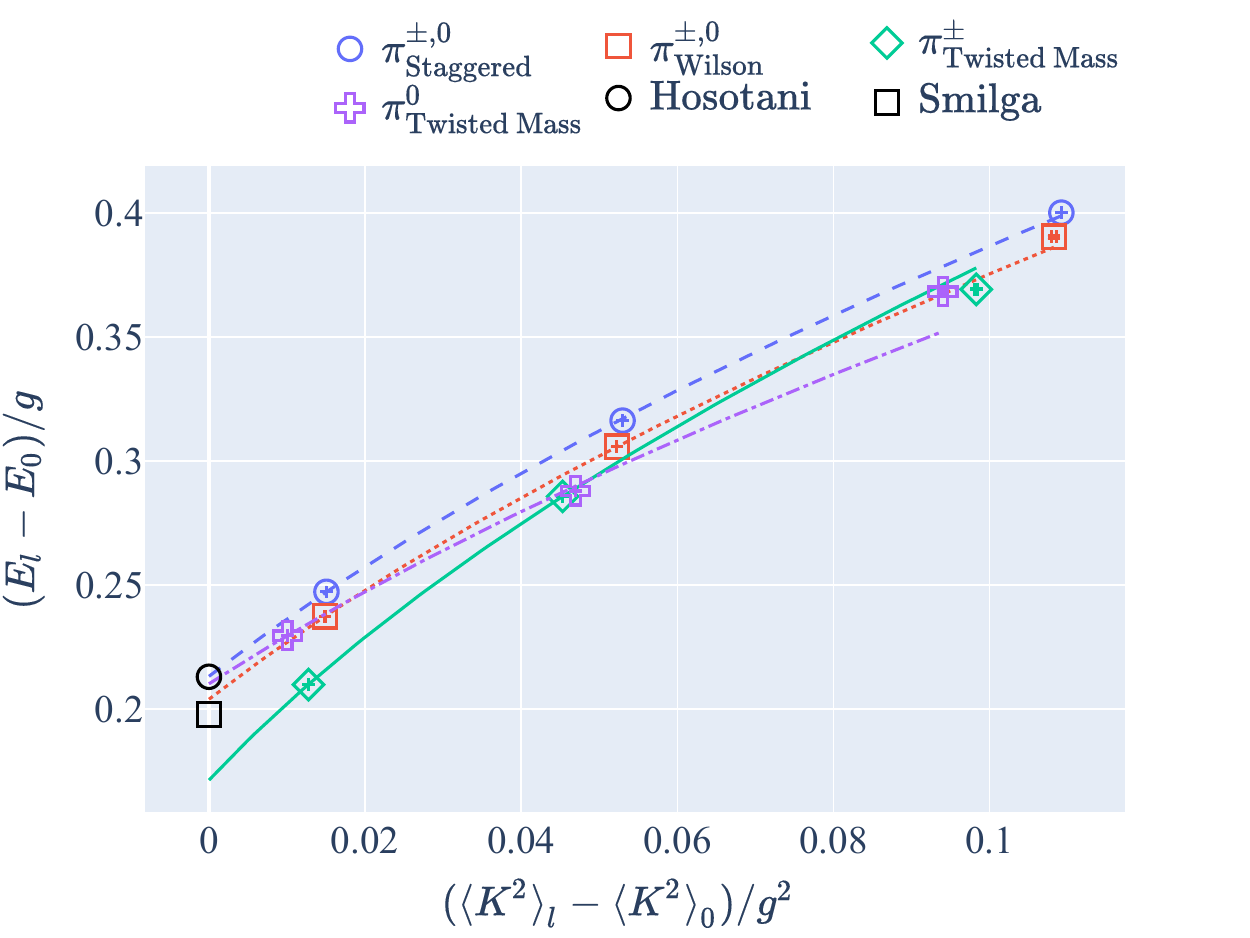}
    \caption{Dispersion fits at finite lattice spacing for the three fermion discretizations, using excitation energy and pseudo-momentum estimates. Staggered and Wilson fermions yield degenerate pion states, while twisted mass fermions show isospin splitting due to lattice artifacts. Extrapolations to zero momentum are compared to analytical predictions for the pion mass from Refs.~\cite{Smilga1997, Hosotani1998}. 
    }
\label{fig:dispersion}
\end{figure}

The resulting dispersion fits for the three fermion discretizations are shown in Fig.~\ref{fig:dispersion}, and the corresponding fit parameters are summarized in Tab.~\ref{tab:dispersion}.

\begin{table}[ht]
  \centering
  \renewcommand{\arraystretch}{1.2} 
  \setlength{\tabcolsep}{10pt}       
  \begin{tabular}{llccc}
    \toprule
    Fermion Type & Pion & \(b\) & \(m_{\pi}/g\) \\
    \midrule
    \multirow{1}{*}{Staggered} 
        & \(\pi^{\pm,0}\) & 1.02 & 0.213 \\
    \midrule
    \multirow{1}{*}{Wilson} 
        & \(\pi^{\pm,0}\) & 0.997 & 0.204 \\
    \midrule
    \multirow{2}{*}{Twisted Mass} 
        & \(\pi^{\pm}\) & 1.07 & 0.171 \\
        & \(\pi^{0}\) & 0.922 & 0.210 \\
    \bottomrule
  \end{tabular}
   \caption{Fit parameters from the dispersion relations in Fig.~\ref{fig:dispersion}. The parameter \( b \) quantifies lattice-induced deviations from the continuum dispersion, while \( m_{\pi}/g \) gives the pion mass at finite lattice spacing.}
  \label{tab:dispersion}
\end{table}

For staggered and Wilson fermions, the extracted masses of the charged and neutral pion are degenerate within numerical precision. The corresponding dispersion relations closely follow the expected continuum form. The pion mass obtained from staggered fermions shows excellent agreement with the analytical prediction of Ref.~\cite{Hosotani1998}, while the result for Wilson fermions is slightly smaller but remains in close proximity to both the analytical value and the staggered result.

In the case of twisted mass fermions, the neutral pion agrees well with both the staggered result and the theoretical prediction. Its dispersion relation is noticeably flatter, consistent with a reduced pseudo-momentum contribution. In contrast, the charged pion exhibits a significantly steeper dispersion curve, leading to a substantial underestimation of the pion mass. This suggests that the dominant finite-volume effects in this channel cannot be adequately described by a simple kinetic correction.

Interestingly, the energy gap of the lowest-momentum charged pion excitation lies remarkably close to the pion mass values extracted from dispersion fits across all three fermion discretizations. While this agreement may be coincidental, it could also hint at a deeper physical mechanism, meriting further investigation.

\subsection{\label{sec:pion_finite_size}Pion mass from finite-volume scaling}

Ultimately, the momentum correction observed in the previous analysis arises from simulating the system in a large but finite volume of \( Lg = 30 \). In the infinite-volume limit, an excitation representing a stable particle with a well-defined mass must be at rest, i.e., carry zero momentum. Finite-volume scaling thus provides an effective approach for extracting pion masses directly from the energy gap. This method accounts for all finite-volume effects---not only momentum corrections---and does not rely on any specific assumption about the dispersion relation. While it requires simulations at multiple volumes, it has the advantage that only the ground state and one excited state need to be computed.

For the simpler one-flavor Schwinger model, this finite-volume scaling approach has already been successfully applied in Ref.~\cite{Banuls2013}. In our study, we consider volumes \( Lg = 20, 25, 30, 35 \). To keep the lattice spacing fixed across these volumes, we adjust the number of lattice sites \( N \) according to the relation \( Lg = N / \sqrt{x} \), where \( x \) is chosen based on the reference setup with \( Lg = 30 \) and \( N = 110 \). This yields a fixed lattice spacing of \( 1/\sqrt{x} = 0.273 \), which is also used in the dispersion relation analysis presented in the previous section.

The renormalized fermion mass is again set to \( m_{\mathrm{r}}/g = 0.1 \), consistent with the dispersion-based study of the pion mass. We reuse the mass renormalization obtained in the initial volume of \( Lg = 30 \), justified by the observation in Ref.~\cite{Angelides2025} for the one-flavor case that the volume dependence of the mass shift plateaus for \( Lg \gtrsim 30 \).

In Ref.~\cite{Hamer1997a}, it was observed that finite-volume effects in the Schwinger model with open boundary conditions exhibit a polynomial scaling behavior. It was argued that the leading finite-volume corrections arise from a kinetic energy contribution of order \( \mathcal{O}(1/(Lg)^2) \), in contrast to the exponential suppression typically expected for periodic boundary conditions. We therefore follow the approach of Ref.~\cite{Banuls2013} and fit the energy gaps using the functional form
\begin{equation}
\label{eq:poly_fv_fit}
    \Delta E = m_{\pi} + \frac{A}{(Lg)^2} + \mathcal{O}\left(\frac{1}{(Lg)^3}\right),
\end{equation}
with fit parameters \( A \) and \( m_{\pi} \). This ansatz accurately captures the finite-volume dependence for staggered and Wilson fermions, as well as for the neutral pion in the twisted mass formulation. We also apply this form to model the volume dependence of the pseudo-momentum for all three fermion discretizations.

For the charged pion in the twisted mass formulation, however, this polynomial form leads to inconsistent extrapolations compared to the other discretizations, given the range of volumes considered. Furthermore, the failure of the dispersion relation approach for the charged pion in Sec.~\ref{sec:pion_mass_dispersion} suggests that the dominant finite-volume effects in this case cannot be attributed solely to kinetic corrections.

Instead, we employ the universal form of finite-volume corrections in massive quantum field theories derived for periodic boundary conditions~\cite{Luscher1985}, fitting the charged pion mass using
\begin{equation}
\label{eq:exp_fv_fit}
    \Delta E = m_{\pi} + B \frac{\sqrt{m_{\pi}}}{\sqrt{Lg}} e^{-L m_{\pi}},
\end{equation}
with fit parameters \( B \) and \(m_{\pi}\). We find that this functional form describes the data for the charged pion in the twisted mass formulation well and yields results consistent with those of the other discretizations. In contrast, it does not provide an accurate description of the volume dependence for staggered and Wilson fermions, which remain better captured by the polynomial ansatz.

\begin{figure*}[ht] 
    \centering
    \includegraphics[width=1.0\textwidth]{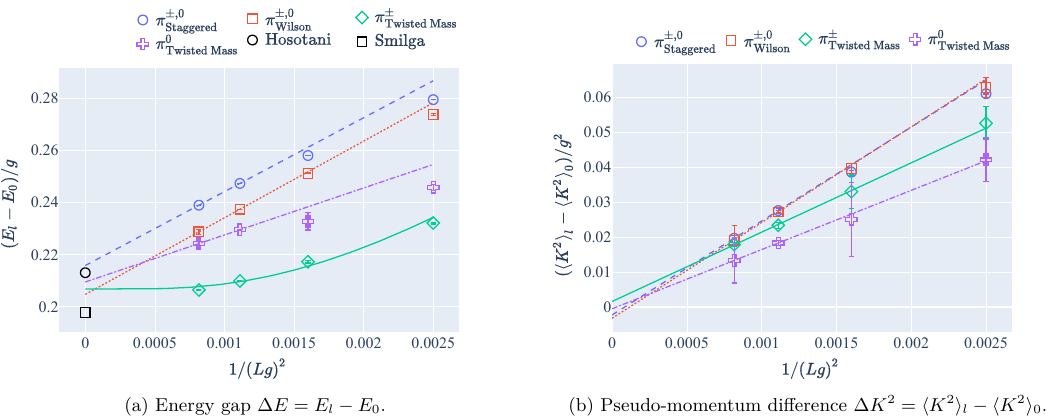}
    \caption{Finite-volume scaling of energy gaps (a) and pseudo-momentum differences (b) at fixed renormalized mass \( m_{\mathrm{r}}/g = 0.1 \). Lines show fits in \(1/Lg\): a quadratic fit is used for staggered and Wilson fermions and for the neutral pion in the twisted mass formulation, while an exponential fit is applied for the charged pion in the twisted mass case. The infinite-volume extrapolations are compared to analytical predictions for the pion mass from Refs.~\cite{Smilga1997, Hosotani1998}. The momentum gap extrapolates to zero, as expected.}
    \label{fig:finite_size}
\end{figure*}

The resulting fits for the three fermion discretizations are shown in Fig.~\ref{fig:finite_size}, and the corresponding fit parameters are summarized in Tab.~\ref{tab:finite_size}.

\begin{table}[ht]
  \centering
  \renewcommand{\arraystretch}{1.2} 
  \setlength{\tabcolsep}{10pt}       
  \begin{tabular}{llcccc}
    \toprule
    Fermion Type & Pion & \(A\) & \(B\) & \(m_{\pi}/g\) \\
    \midrule
    \multirow{1}{*}{Staggered} 
        & \(\pi^{\pm,0}\) & 28.4 & & 0.216 \\
    \midrule
    \multirow{1}{*}{Wilson} 
        & \(\pi^{\pm,0}\) & 29.4 & & 0.205 \\
    \midrule
    \multirow{2}{*}{Twisted Mass} 
        & \(\pi^{\pm}\) &  & 16.9 & 0.207 \\
        & \(\pi^{0}\) & 18.0 & & 0.209 \\
    \bottomrule
  \end{tabular}
\caption{Fit parameters from the finite-volume scaling analysis in Fig.~\ref{fig:finite_size}. The coefficient \( A \) captures the leading volume dependence, while \( m_{\pi}/g \) denotes the pion mass at finite lattice spacing.}
  \label{tab:finite_size}
\end{table}

The study of the volume dependence of the excitation gaps reveals several noteworthy features. For sufficiently large volumes (\( Lg \gtrsim 25 \)), the data is well described by the expected quadratic scaling in \( 1/(Lg) \). At smaller volumes, however, higher-order corrections become visible, leading to deviations from the quadratic behavior. The magnitude of these deviations depends on the fermion discretization: they are more pronounced for staggered fermions and for the neutral pion in the twisted mass formulation, while they remain comparatively mild for Wilson fermions. In contrast, the charged pion in the twisted mass formulation is well described by the universal exponential form derived for massive quantum field theories, which accurately captures its volume dependence.

The pseudo-momentum differences follow the expected quadratic scaling with good accuracy and extrapolate to zero in the infinite-volume limit, as anticipated. Overall, all pion masses obtained from the three fermion discretizations differ by at most \( 5\% \), even at the fixed lattice spacing of \( 1/\sqrt{x} = 0.273 \).

In a final step, we combine finite-volume scaling with a continuum extrapolation. We perform this computation for staggered and Wilson fermions, as well as for the charged pion in the twisted mass formulation. The neutral pion in twisted mass fermions is excluded from this analysis, as it requires access to higher excited states. We note, however, that both pion channels in the twisted mass formulation are expected to yield the same continuum limit, since isospin breaking effects vanish as \( a \to 0 \).

At each fixed lattice spacing, we simulate at volumes \( Lg = 20, 25, 30, 35 \). For staggered and Wilson fermions, the infinite-volume value is extracted via a linear fit in \(1/(Lg)^2\) using the two largest volumes. The associated uncertainty is estimated using the jackknife resampling procedure described in Sec.~\ref{sec:electric_field_density}, applied across all volumes to account for potential nonlinearities. This analysis employs the polynomial form introduced in Eq.~\eqref{eq:poly_fv_fit}.

For the charged pion in the twisted mass formulation, we use the exponential ansatz from Eq.~\eqref{eq:exp_fv_fit}, fitting all available volumes and estimating uncertainties via the same jackknife procedure.

\begin{figure*}[ht]
    \centering
    \includegraphics[width=1.0\textwidth]{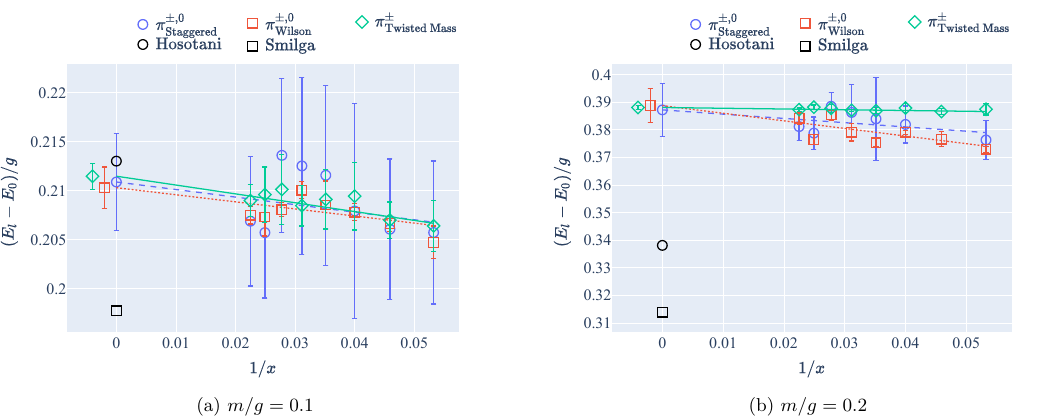}
    \caption{Combined continuum and finite-volume extrapolation of the pion mass for staggered, Wilson, and twisted mass fermions, including mass renormalization. Each data point represents a volume-extrapolated result at fixed lattice spacing, obtained from simulations at \( Lg = 20\text{--}35 \). The continuum limit is extracted via a purely quadratic fit in \(1/\sqrt{x}\). Analytical predictions from Refs.~\cite{Smilga1997, Hosotani1998} are shown for comparison.}
    \label{fig:mass_gap_continuum_finite_size}
\end{figure*}

Following this finite-volume analysis, we perform a purely quadratic extrapolation to the continuum limit, with uncertainties again obtained using the jackknife method. The resulting fits are shown in Fig.~\ref{fig:mass_gap_continuum_finite_size}. For comparison, we include analytical predictions. We find excellent agreement across all lattice data within uncertainties.

\begin{figure}
    \centering
    \includegraphics[width=\linewidth]{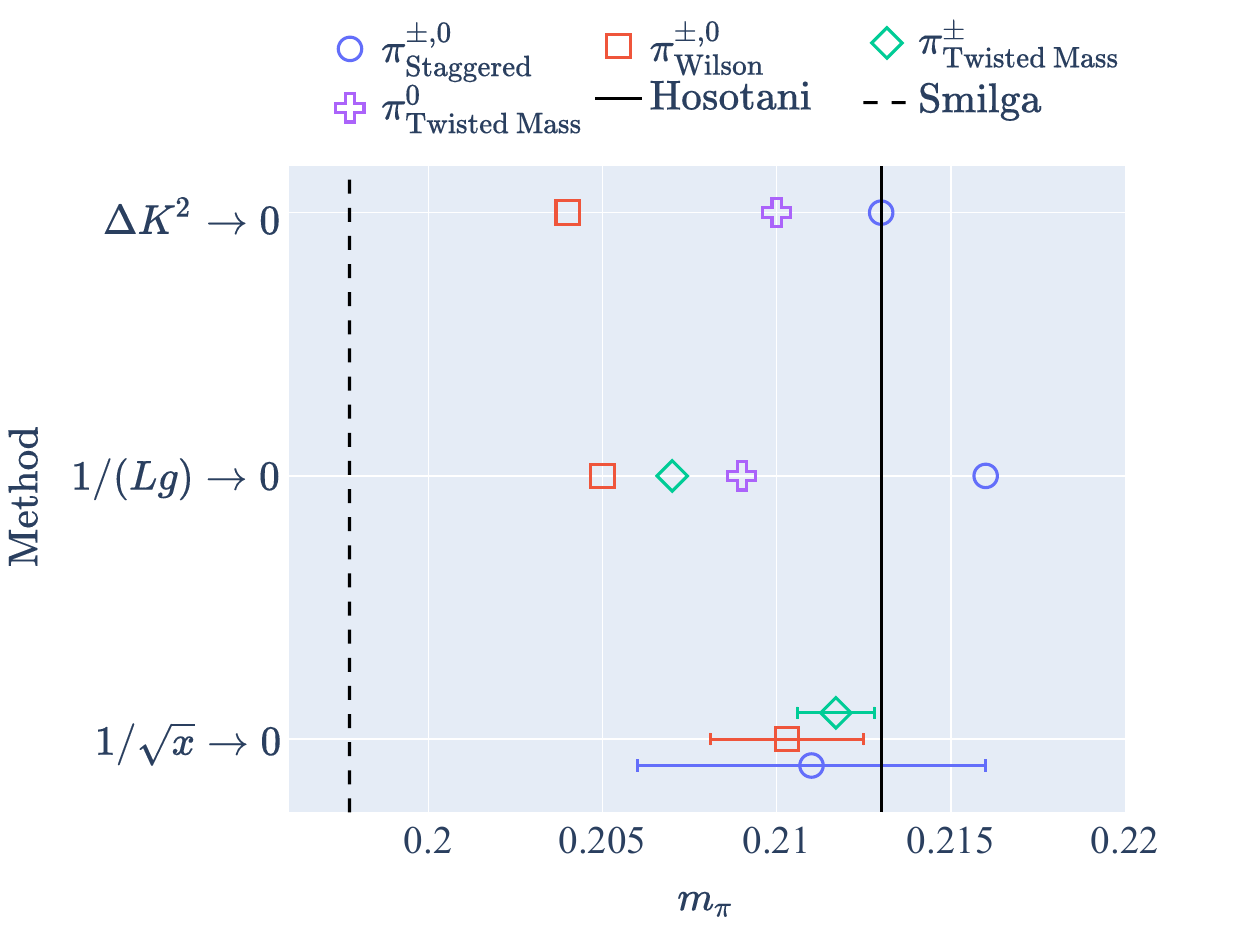}
    \caption{Pion mass results from various fermion discretizations and extraction methods for fermion mass \( m/g = 0.1 \). Shown (from top to bottom) are: results from dispersion relation fits at fixed lattice spacing \( 1/\sqrt{x} = 0.273 \) and fixed volume \( Lg = 30 \); finite-volume scaling results at the same lattice spacing; and continuum extrapolations performed in the thermodynamic limit. Horizontal lines indicate analytical predictions from Refs.~\cite{Hosotani1998,Smilga1997}. The charged pion mass from the dispersion relation for twisted mass fermions lies well below the plot range at \( m_{\pi}/g = 0.171 \).}
    \label{fig:pion_mass_overview}
\end{figure}

We conclude with an overview of our results for the pion mass, as summarized in Fig.~\ref{fig:pion_mass_overview} and Tab.~\ref{tab:pion_mass_overview}. We compare continuum extrapolations performed in the thermodynamic limit with two complementary extraction methods applied at fixed lattice spacing \( 1/\sqrt{x} = 0.273 \): the dispersion relation approach (\( \Delta K^2 \to 0 \)) and finite-volume scaling (\( 1/(Lg) \to 0 \)). These comparisons provide a comprehensive consistency check across discretizations and methods, and allow us to assess both discretization and finite-volume effects.

\begin{table}[ht]
  \centering
  \renewcommand{\arraystretch}{1.2} 
  \setlength{\tabcolsep}{8pt}       
  \begin{tabular}{lcc}
    \toprule
    Result & \({m}/{g}=0.1\) & \({m}/{g}=0.2\) \\
    \midrule
    Staggered & 0.211(5) & 0.386(11) \\
    \midrule
    Wilson & 0.2103(22) & 0.389(7) \\
    \midrule
    Twisted Mass & 0.2117(11) & 0.3880(8) \\
    \midrule
    Hosotani & 0.213 & 0.338 \\
    Smilga & 0.198 & 0.314 \\
    \bottomrule
  \end{tabular}
  \caption{Continuum-extrapolated pion mass for staggered, Wilson and twisted mass fermions, compared to analytical predictions from Refs.~\cite{Hosotani1998,Smilga1997}. Results are shown for two values of the fermion mass, \( m/g = 0.1 \) and \( m/g = 0.2 \). The data show excellent agreement with the Hosotani prediction at \( m/g = 0.1 \), while the Smilga estimate lies systematically below the numerical results.}
  \label{tab:pion_mass_overview}
\end{table}

We observe excellent agreement between the continuum-extrapolated pion masses for the three fermion discretizations, with all results compatible within their respective uncertainties. These continuum values incorporate all sources of uncertainty addressed in our study: numerical errors due to finite bond dimension, systematic uncertainties from finite-volume extrapolation, and statistical uncertainties from the continuum limit fits. We therefore regard these values as our most precise results and use them as reference benchmarks.

The comparatively larger uncertainty associated with staggered fermions is primarily due to their more pronounced sensitivity to finite-volume effects in the parameter regime explored. These uncertainties could be further reduced by performing simulations at larger physical volumes. However, since the primary aim of this study was to assess and compare the scaling behavior of different fermion discretizations under identical lattice parameters, we did not individually optimize simulation volumes for each formulation.

At finite lattice spacing, differences between fermion discretizations dominate over the specific method used to account for finite-volume effects—whether via dispersion relation fits or finite-size scaling. For staggered fermions and the neutral pion in the twisted mass formulation, the deviation from the continuum limit remains below approximately \(1\%\). In contrast, Wilson fermions yield systematically lower values with deviations around \(3\%\). The finite-size scaling result for the charged pion in the twisted mass formulation lies slightly below its neutral counterpart due to isospin breaking effects. Notably, the dispersion relation result for the charged twisted mass pion lies significantly below all other values, indicating that finite-volume effects in this channel are not well captured by a simple kinetic correction.

Our numerical results show excellent agreement with the analytical prediction of Ref.~\cite{Hosotani1998}, while the estimate from Ref.~\cite{Smilga1997} lies significantly below all lattice results, suggesting that it underestimates the pion mass in the intermediate mass regime explored here.

\section{\label{sec:discussion}Discussion and outlook}

In this paper, we have presented extensive numerical studies of staggered, Wilson, and for the first time of twisted mass fermions within the Hamiltonian formalism. As a benchmark, we employed the massive two-flavor Schwinger model—a physically rich yet computationally accessible setting. The uncertainties introduced by the here employed tensor network method were controlled to a level that enabled precise and statistically significant results.

A particular focus of this work was the investigation of twisted mass fermions, which extend the set of fermion discretizations commonly used in Hamiltonian-based TNS simulations. While most previous studies have relied on staggered fermions due to their simplicity and efficiency, and only a few have explored Wilson fermions, this work constitutes the first systematic study of twisted mass fermions in this context. As a preliminary check, we confirmed the expected \(\mathcal{O}(a)\) improvement at maximal twist in the free theory, providing initial evidence that this may also hold in the interacting case.

Analyzing the scaling behavior of different fermion discretizations is challenging, since it turned out that lattice artifacts are small and the computational cost of accessing sufficiently fine lattice spacings is high. By increasing the fermion mass—thereby effectively tuning the theory closer to maximal twist—we were able to observe automatic \( \mathcal{O}(a) \) improvement for twisted mass fermions, using the electric field density as a representative observable. In contrast, staggered and Wilson fermions exhibited predominantly linear scaling behavior.

Standard techniques for computing the mass renormalization in the Lagrangian formalism do not directly carry over to the Hamiltonian setting. We therefore adopted an alternative method based on identifying the zero crossing of the electric field density. This required scanning small negative values of the lattice mass, where the system becomes nearly gapless and numerically challenging. Nevertheless, by enforcing appropriate quantum number constraints, we were able to reliably extract the ground state in this regime.

Our results confirm that this electric-field-based method is applicable to the two-flavor model and is suitable for tuning twisted mass fermions to maximal twist. Moreover, we verified the analytical prediction for the additive mass renormalization derived under periodic boundary conditions, using simulations with open boundary conditions.

We then proceeded to study the pion mass. Once mass renormalization was included, we observed rapid convergence of the excitation energy to the continuum limit, with finite-volume effects becoming the dominant source of systematic uncertainty. To correct for these, we applied two complementary approaches: one based on the dispersion relation and one based on finite-volume scaling. Both yielded consistent results for the pion mass and allowed us to compare and relate the methods.

We compared our numerical results to analytical predictions available in the literature. Remarkably, we found excellent agreement with the estimate by Ref.~\cite{Hosotani1998}, which is based on a semiclassical Hartree-Fock treatment of the model. In contrast, our results showed a modest but systematic deviation from the prediction by Ref.~\cite{Smilga1997}, derived via effective field theory arguments and matched to universal amplitude ratios obtained from the thermodynamic Bethe ansatz. Since our simulations were performed at an intermediate fermion mass of \( m/g \sim 0.1 \), this suggests that the semiclassical approach better captures the physics in this parameter regime. We expect the effective field theory result to become increasingly accurate in the deep infrared, i.e., for asymptotically small fermion masses beyond the range explored here.

Importantly, we found that twisted mass fermions exhibit significantly milder volume dependence compared to staggered and Wilson fermions, making them especially attractive for simulations at moderate volumes. In addition, we clearly resolved isospin breaking effects in the twisted mass formulation—well known from lattice QCD. 

In computing the mass renormalization, we also encountered an intriguing phase structure, reminiscent of lattice QCD, where multiple mass renormalization schemes—particularly in the context of twisted mass fermions—are well established. Exploring analogous structures within the Hamiltonian formalism would be an interesting direction for future work. In addition, the electric-field-density method for determining the mass shift warrants further study. In particular, it would be valuable to understand how the method's intrinsic uncertainties propagate to final observables. This could reveal both limitations and potential strengths and may motivate alternative strategies, for instance based on entanglement entropy, the mass gap \cite{guo2024}, or other order parameters.

In this work, we have carried out a combined continuum and finite-volume extrapolation for all three fermion discretizations—staggered, Wilson, and twisted mass fermions—yielding precise and consistent results for the pion mass in the two-flavor Schwinger model. These results provide a reliable benchmark for future studies and highlight the capabilities of tensor network methods in extracting physical observables with controlled uncertainties. Looking ahead, the main challenge remains extending these techniques to higher dimensions, ultimately aiming toward simulations of lattice QCD. In this context, our findings underscore the potential of twisted mass fermions as a promising discretization choice: they fully resolve the doubling problem in \(3+1\) dimensions—unlike staggered fermions—and exhibit improved scaling behavior compared to standard Wilson fermions.

\begin{acknowledgments}
    We thank Maria-Eftaxia Stasinou und Kostas Blekos for helpful discussions. 
    This work is supported with funds from the Ministry of Science, Research and Culture of the State of Brandenburg within the Center for Quantum Technology and Applications (CQTA). 
    \begin{center}
        \includegraphics[width = 0.08\textwidth]{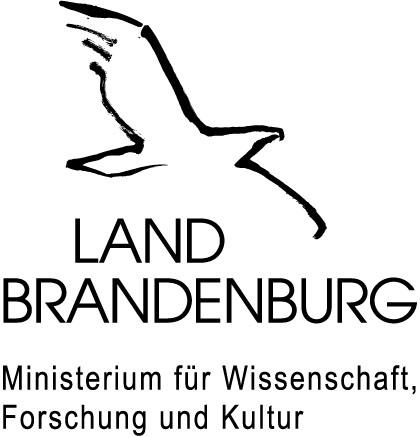}
    \end{center}
\end{acknowledgments}

\appendix

\section{Free Fermion Spectrum for Wilson and Twisted Mass Fermions}
\label{sec:free_fermion_spectrum}

The lattice theory of free fermions can be solved exactly. For staggered fermions, this was done in Ref.~\cite{Rigobello2021} for periodic boundary conditions and in Ref.~\cite{Banuls2016} for open boundary conditions. Here, we demonstrate how to obtain the spectrum of free Wilson and twisted mass fermions, extending the approach to the two-flavor case.

\subsection{Hamiltonian and Conventions}

The Dirac Hamiltonian for two flavors of free Wilson (twisted mass) fermions with periodic boundary conditions is given by
\begin{equation}
\begin{split}
H_D &= \sum_{n=0}^{N-1} \sum_{f=0}^{1} a\, \psi_{n,f}^{\dag} \gamma^0 \left( m + (-1)^f i \mu \gamma^5 + \frac{r}{a} \right) \psi_{n,f} \\
&\quad - \frac{1}{2} \sum_{n=0}^{N-1} \sum_{f=0}^{1} \left( \psi_{n,f}^{\dag} \gamma^0 (i\gamma^1 + r)\psi_{n+1,f} + \text{h.c.} \right),
\end{split}
\end{equation}
where \(f = 0, 1\) indexes the two fermion flavors, and the twisted mass term introduces a relative sign between them.

We adopt the conventions
\begin{equation}
r = 1, \quad \gamma^0 = X, \quad \gamma^1 = iZ, \quad \gamma^5 = \gamma^0 \gamma^1,
\end{equation}
which imply
\begin{equation}
i\gamma^0\gamma^1 + r\gamma^0 = 
\begin{pmatrix}
0 & 2 \\
0 & 0  
\end{pmatrix}, \quad
i\gamma^0\gamma^5 = i\gamma^1 = 
\begin{pmatrix}
-1 & 0 \\ 
0 & 1 
\end{pmatrix}.
\end{equation}

We rescale the fields via \(\phi_{n,f} = \sqrt{a} \psi_{n,f}\), eliminating factors of \(a\) from the kinetic terms and ensuring the fields are dimensionless. This also simplifies the normalization of the Fourier transform and delta functions.

The Hamiltonian in terms of components becomes
\begin{equation}
\begin{split}
H_D &= \sum_{n=0}^{N-1} \sum_{f=0}^{1} \left( m + \frac{1}{a} \right) \left( \phi_{n,f,1}^{\dag}\phi_{n,f,2} + \text{h.c.} \right) \\
&\quad - \sum_{n=0}^{N-1} \sum_{f=0}^{1} (-1)^f \mu \left( \phi_{n,f,1}^{\dag}\phi_{n,f,1} - \phi_{n,f,2}^{\dag}\phi_{n,f,2} \right) \\
&\quad - \frac{1}{a} \sum_{n=0}^{N-1} \sum_{f=0}^{1} \left( \phi_{n,f,1}^{\dag}\phi_{n+1,f,2} + \text{h.c.} \right).
\end{split}
\end{equation}

\subsection{Fourier Transform and Momentum-Space Hamiltonian}

We apply the discrete Fourier transform:
\begin{equation}
\begin{split}
\phi_{k,f,\alpha} &= \frac{1}{\sqrt{N}} \sum_{n=0}^{N-1} e^{-i2\pi kn/N} \phi_{n,f,\alpha} \\
\phi_{n,f,\alpha} &= \frac{1}{\sqrt{N}} \sum_{k=0}^{N-1} e^{i2\pi kn/N} \phi_{k,f,\alpha}.
\end{split}
\end{equation}

Using the identity
\begin{equation}
\sum_{n=0}^{N-1} e^{i2\pi(k-k')n/N} = N \delta_{kk'},
\end{equation}
we find
\begin{equation}
\sum_{n=0}^{N-1} \phi_{n,f,\alpha}^{\dag}\phi_{n,f',\alpha'} = \sum_{k=0}^{N-1} \phi_{k,f,\alpha}^{\dag}\phi_{k,f',\alpha'},
\end{equation}
and similarly,
\begin{equation}
\sum_{n=0}^{N-1} \phi_{n,f,\alpha}^{\dag}\phi_{n+1,f',\alpha'} = \sum_{k=0}^{N-1} e^{i2\pi k/N} \phi_{k,f,\alpha}^{\dag}\phi_{k,f',\alpha'}.
\end{equation}

The Hamiltonian in momentum space becomes
\begin{equation}
\begin{split}
H_D &= \sum_{k=0}^{N-1} \sum_{f=0}^{1} \Bigg[ \left( m + \frac{1}{a} \right) \left( \phi_{k,f,1}^{\dag}\phi_{k,f,2} + \text{h.c.} \right) \\
&\quad - (-1)^f \mu \left( \phi_{k,f,1}^{\dag}\phi_{k,f,1} - \phi_{k,f,2}^{\dag}\phi_{k,f,2} \right) \\
&\quad - \frac{1}{a} \left( e^{i2\pi k/N} \phi_{k,f,1}^{\dag}\phi_{k,f,2} + \text{h.c.} \right) \Bigg].
\end{split}
\end{equation}

\subsection{Matrix Form and Spectrum}

We express the Hamiltonian in matrix form:
\begin{equation}
\begin{split}
H_D &= \sum_{k=0}^{N-1} \sum_{f=0}^{1} H_{k,f}, \\
H_{k,f} &= 
\begin{pmatrix}
\phi_{k,f,1}^{\dag} \\
\phi_{k,f,2}^{\dag}
\end{pmatrix}^{\!\mathsf{T}}
\begin{pmatrix}
-(-1)^f \mu & \Delta_k \\
\Delta_k^* & (-1)^f \mu
\end{pmatrix}
\begin{pmatrix}
\phi_{k,f,1} \\
\phi_{k,f,2}
\end{pmatrix}, \\
\Delta_k &= m + \frac{1}{a} \left(1 - e^{i2\pi k/N} \right).
\end{split}
\end{equation}

In the case of standard Wilson fermions (\(\mu = 0\)), the diagonal entries vanish. The eigenvalues are given by:
\begin{equation}
\lambda_\pm = \pm \sqrt{A^2 + |B|^2}, \quad \text{with } A = (-1)^f \mu, \quad B = \Delta_k.
\end{equation}

The resulting expression gives the energy dispersion of free Wilson (twisted mass) fermions:
\begin{equation}
\begin{split}
\lambda_+ &= \sqrt{ \mu^2 + \left( m + \frac{1}{a}(1 - \cos(2\pi k/N)) \right)^2 + \frac{1}{a^2} \sin^2(2\pi k/N) } \\
&= \sqrt{ \mu^2 + m^2 + \frac{2}{a}(m + \frac{1}{a})(1 - \cos(2\pi k/N)) }.
\end{split}
\end{equation}

The ground state energy is given by
\begin{equation}
E_0 = -2 \sum_{k=0}^{N-1} \lambda_+,
\end{equation}
where the factor of 2 accounts for the two fermion flavors, each filling the negative-energy states in the Dirac sea.

\subsection{Continuum Expansion}

Expanding in the lattice spacing \(a\) and defining the physical momentum \(k' = 2\pi k / L\), we obtain:
\begin{equation}
\lambda_+ = \sqrt{C} + \frac{mk'^2}{2\sqrt{C}}\, a - \frac{3m^2k'^4 + Ck'^4}{12\, C^{3/2}}\, a^2 + \mathcal{O}(a^3),
\end{equation}
where \(C = \mu^2 + m^2 + k'^2\).

We observe that the \(\mathcal{O}(a)\) term vanishes for \(m = 0\), in line with the expected automatic \(\mathcal{O}(a)\) improvement of twisted mass fermions at maximal twist.

\bibliography{references}

\begin{thebibliography}{68}%
\makeatletter
\providecommand \@ifxundefined [1]{%
 \@ifx{#1\undefined}
}%
\providecommand \@ifnum [1]{%
 \ifnum #1\expandafter \@firstoftwo
 \else \expandafter \@secondoftwo
 \fi
}%
\providecommand \@ifx [1]{%
 \ifx #1\expandafter \@firstoftwo
 \else \expandafter \@secondoftwo
 \fi
}%
\providecommand \natexlab [1]{#1}%
\providecommand \enquote  [1]{``#1''}%
\providecommand \bibnamefont  [1]{#1}%
\providecommand \bibfnamefont [1]{#1}%
\providecommand \citenamefont [1]{#1}%
\providecommand \href@noop [0]{\@secondoftwo}%
\providecommand \href [0]{\begingroup \@sanitize@url \@href}%
\providecommand \@href[1]{\@@startlink{#1}\@@href}%
\providecommand \@@href[1]{\endgroup#1\@@endlink}%
\providecommand \@sanitize@url [0]{\catcode `\\12\catcode `\$12\catcode `\&12\catcode `\#12\catcode `\^12\catcode `\_12\catcode `\%12\relax}%
\providecommand \@@startlink[1]{}%
\providecommand \@@endlink[0]{}%
\providecommand \url  [0]{\begingroup\@sanitize@url \@url }%
\providecommand \@url [1]{\endgroup\@href {#1}{\urlprefix }}%
\providecommand \urlprefix  [0]{URL }%
\providecommand \Eprint [0]{\href }%
\providecommand \doibase [0]{https://doi.org/}%
\providecommand \selectlanguage [0]{\@gobble}%
\providecommand \bibinfo  [0]{\@secondoftwo}%
\providecommand \bibfield  [0]{\@secondoftwo}%
\providecommand \translation [1]{[#1]}%
\providecommand \BibitemOpen [0]{}%
\providecommand \bibitemStop [0]{}%
\providecommand \bibitemNoStop [0]{.\EOS\space}%
\providecommand \EOS [0]{\spacefactor3000\relax}%
\providecommand \BibitemShut  [1]{\csname bibitem#1\endcsname}%
\let\auto@bib@innerbib\@empty
\bibitem [{\citenamefont {Gattringer}\ and\ \citenamefont {Lang}(2010)}]{Gattringer2010}%
  \BibitemOpen
  \bibfield  {author} {\bibinfo {author} {\bibfnamefont {C.}~\bibnamefont {Gattringer}}\ and\ \bibinfo {author} {\bibfnamefont {C.~B.}\ \bibnamefont {Lang}},\ }\href {https://doi.org/10.1007/978-3-642-01850-3} {\emph {\bibinfo {title} {Quantum chromodynamics on the lattice}}},\ Vol.\ \bibinfo {volume} {788}\ (\bibinfo  {publisher} {Springer},\ \bibinfo {address} {Berlin},\ \bibinfo {year} {2010})\BibitemShut {NoStop}%
\bibitem [{\citenamefont {Rothe}(2012)}]{Rothe2012}%
  \BibitemOpen
  \bibfield  {author} {\bibinfo {author} {\bibfnamefont {H.~J.}\ \bibnamefont {Rothe}},\ }\href {https://doi.org/10.1142/8229} {{\selectlanguage {English}\emph {\bibinfo {title} {Lattice {Gauge} {Theories}: {An} {Introduction} ({Fourth} {Edition})}}}}\ (\bibinfo  {publisher} {World Scientific Publishing Company},\ \bibinfo {year} {2012})\BibitemShut {NoStop}%
\bibitem [{\citenamefont {{Y.~Aoki and T.~Blum and S.~Collins and L.~Del\.{}Debbio and M.~Della\,Morte and P.~Dimopoulos and X.~Feng and M.~Golterman and Steven~Gottlieb and R.~Gupta and G.~Herdoiza and P.~Hernandez and A.~J\"{u}ttner and T.~Kaneko and E.~Lunghi and S.~Meinel and C.~Monahan and A.~Nicholson and T.~Onogi and P.~Petreczky and A.~Portelli and A.~Ramos and S.R.~Sharpe and J.N.~Simone and S.~Sint and R.~Sommer and N.~Tantalo and R.~Van\,de\,Water and A.~Vaquero and U.~Wenger and H.~Wittig}}(202)}]{FLAG2024}%
  \BibitemOpen
  \bibfield  {author} {\bibinfo {author} {\bibnamefont {{Y.~Aoki and T.~Blum and S.~Collins and L.~Del\.{}Debbio and M.~Della\,Morte and P.~Dimopoulos and X.~Feng and M.~Golterman and Steven~Gottlieb and R.~Gupta and G.~Herdoiza and P.~Hernandez and A.~J\"{u}ttner and T.~Kaneko and E.~Lunghi and S.~Meinel and C.~Monahan and A.~Nicholson and T.~Onogi and P.~Petreczky and A.~Portelli and A.~Ramos and S.R.~Sharpe and J.N.~Simone and S.~Sint and R.~Sommer and N.~Tantalo and R.~Van\,de\,Water and A.~Vaquero and U.~Wenger and H.~Wittig}}},\ }\bibfield  {title} {\bibinfo {title} {{FLAG Review 2024}},\ }\href@noop {} {\bibfield  {journal} {\bibinfo  {journal} {arXiv preprint}\ } (\bibinfo {year} {2024})},\ \bibinfo {note} {[v2, last revised 17 Jan 2025]},\ \Eprint {https://arxiv.org/abs/2411.04268} {arXiv:2411.04268 [hep-lat]} \BibitemShut {NoStop}%
\bibitem [{\citenamefont {Loh}\ \emph {et~al.}(1990)\citenamefont {Loh}, \citenamefont {Gubernatis}, \citenamefont {Scalettar}, \citenamefont {White}, \citenamefont {Scalapino},\ and\ \citenamefont {Sugar}}]{Loh1990}%
  \BibitemOpen
  \bibfield  {author} {\bibinfo {author} {\bibfnamefont {E.~Y.}\ \bibnamefont {Loh}}, \bibinfo {author} {\bibfnamefont {J.~E.}\ \bibnamefont {Gubernatis}}, \bibinfo {author} {\bibfnamefont {R.~T.}\ \bibnamefont {Scalettar}}, \bibinfo {author} {\bibfnamefont {S.~R.}\ \bibnamefont {White}}, \bibinfo {author} {\bibfnamefont {D.~J.}\ \bibnamefont {Scalapino}},\ and\ \bibinfo {author} {\bibfnamefont {R.~L.}\ \bibnamefont {Sugar}},\ }\bibfield  {title} {\bibinfo {title} {Sign problem in the numerical simulation of many-electron systems},\ }\href {https://doi.org/10.1103/PhysRevB.41.9301} {\bibfield  {journal} {\bibinfo  {journal} {Phys. Rev. B}\ }\textbf {\bibinfo {volume} {41}},\ \bibinfo {pages} {9301} (\bibinfo {year} {1990})}\BibitemShut {NoStop}%
\bibitem [{\citenamefont {Kogut}\ and\ \citenamefont {Susskind}(1975)}]{Kogut1975}%
  \BibitemOpen
  \bibfield  {author} {\bibinfo {author} {\bibfnamefont {J.}~\bibnamefont {Kogut}}\ and\ \bibinfo {author} {\bibfnamefont {L.}~\bibnamefont {Susskind}},\ }\bibfield  {title} {\bibinfo {title} {Hamiltonian formulation of {Wilson}'s lattice gauge theories},\ }\href {https://doi.org/10.1103/PhysRevD.11.395} {\bibfield  {journal} {\bibinfo  {journal} {Phys. Rev. D}\ }\textbf {\bibinfo {volume} {11}},\ \bibinfo {pages} {395} (\bibinfo {year} {1975})}\BibitemShut {NoStop}%
\bibitem [{\citenamefont {Bañuls}\ \emph {et~al.}(2019)\citenamefont {Bañuls}, \citenamefont {Cichy}, \citenamefont {Cirac}, \citenamefont {Jansen},\ and\ \citenamefont {Kühn}}]{Banuls2019}%
  \BibitemOpen
  \bibfield  {author} {\bibinfo {author} {\bibfnamefont {M.~C.}\ \bibnamefont {Bañuls}}, \bibinfo {author} {\bibfnamefont {K.}~\bibnamefont {Cichy}}, \bibinfo {author} {\bibfnamefont {J.~I.}\ \bibnamefont {Cirac}}, \bibinfo {author} {\bibfnamefont {K.}~\bibnamefont {Jansen}},\ and\ \bibinfo {author} {\bibfnamefont {S.}~\bibnamefont {Kühn}},\ }\bibfield  {title} {\bibinfo {title} {Tensor {Networks} and their use for {Lattice} {Gauge} {Theories}},\ }\href {https://doi.org/10.22323/1.334.0022} {\bibfield  {journal} {\bibinfo  {journal} {PoS(LATTICE2018)}\ }\textbf {\bibinfo {volume} {LATTICE2018}},\ \bibinfo {pages} {022} (\bibinfo {year} {2019})}\BibitemShut {NoStop}%
\bibitem [{\citenamefont {Ba{\~{n}}uls}\ and\ \citenamefont {Cichy}(2020)}]{Banuls2019a}%
  \BibitemOpen
  \bibfield  {author} {\bibinfo {author} {\bibfnamefont {M.~C.}\ \bibnamefont {Ba{\~{n}}uls}}\ and\ \bibinfo {author} {\bibfnamefont {K.}~\bibnamefont {Cichy}},\ }\bibfield  {title} {\bibinfo {title} {Review on novel methods for lattice gauge theories},\ }\href {https://doi.org/10.1088/1361-6633/ab6311} {\bibfield  {journal} {\bibinfo  {journal} {Rep. Prog. Phys.}\ }\textbf {\bibinfo {volume} {83}},\ \bibinfo {pages} {024401} (\bibinfo {year} {2020})}\BibitemShut {NoStop}%
\bibitem [{\citenamefont {Ba{\~{n}}uls}\ \emph {et~al.}(2020)\citenamefont {Ba{\~{n}}uls}, \citenamefont {Blatt}, \citenamefont {Catani}, \citenamefont {Celi}, \citenamefont {Cirac}, \citenamefont {Dalmonte}, \citenamefont {Fallani}, \citenamefont {Jansen}, \citenamefont {Lewenstein}, \citenamefont {Montangero}, \citenamefont {Muschik}, \citenamefont {Reznik}, \citenamefont {Rico}, \citenamefont {Tagliacozzo}, \citenamefont {Acoleyen}, \citenamefont {Verstraete}, \citenamefont {Wiese}, \citenamefont {Wingate}, \citenamefont {Zakrzewski},\ and\ \citenamefont {Zoller}}]{Banuls2020}%
  \BibitemOpen
  \bibfield  {author} {\bibinfo {author} {\bibfnamefont {M.~C.}\ \bibnamefont {Ba{\~{n}}uls}}, \bibinfo {author} {\bibfnamefont {R.}~\bibnamefont {Blatt}}, \bibinfo {author} {\bibfnamefont {J.}~\bibnamefont {Catani}}, \bibinfo {author} {\bibfnamefont {A.}~\bibnamefont {Celi}}, \bibinfo {author} {\bibfnamefont {J.~I.}\ \bibnamefont {Cirac}}, \bibinfo {author} {\bibfnamefont {M.}~\bibnamefont {Dalmonte}}, \bibinfo {author} {\bibfnamefont {L.}~\bibnamefont {Fallani}}, \bibinfo {author} {\bibfnamefont {K.}~\bibnamefont {Jansen}}, \bibinfo {author} {\bibfnamefont {M.}~\bibnamefont {Lewenstein}}, \bibinfo {author} {\bibfnamefont {S.}~\bibnamefont {Montangero}}, \bibinfo {author} {\bibfnamefont {C.~A.}\ \bibnamefont {Muschik}}, \bibinfo {author} {\bibfnamefont {B.}~\bibnamefont {Reznik}}, \bibinfo {author} {\bibfnamefont {E.}~\bibnamefont {Rico}}, \bibinfo {author} {\bibfnamefont {L.}~\bibnamefont {Tagliacozzo}}, \bibinfo {author} {\bibfnamefont {K.~V.}\ \bibnamefont {Acoleyen}}, \bibinfo {author} {\bibfnamefont
  {F.}~\bibnamefont {Verstraete}}, \bibinfo {author} {\bibfnamefont {U.-J.}\ \bibnamefont {Wiese}}, \bibinfo {author} {\bibfnamefont {M.}~\bibnamefont {Wingate}}, \bibinfo {author} {\bibfnamefont {J.}~\bibnamefont {Zakrzewski}},\ and\ \bibinfo {author} {\bibfnamefont {P.}~\bibnamefont {Zoller}},\ }\bibfield  {title} {\bibinfo {title} {Simulating lattice gauge theories within quantum technologies},\ }\href {https://doi.org/10.1140/epjd/e2020-100571-8} {\bibfield  {journal} {\bibinfo  {journal} {The European Physical Journal D}\ }\textbf {\bibinfo {volume} {74}},\ \bibinfo {pages} {165} (\bibinfo {year} {2020})}\BibitemShut {NoStop}%
\bibitem [{\citenamefont {Rigobello}\ \emph {et~al.}(2021)\citenamefont {Rigobello}, \citenamefont {Notarnicola}, \citenamefont {Magnifico},\ and\ \citenamefont {Montangero}}]{Rigobello2021}%
  \BibitemOpen
  \bibfield  {author} {\bibinfo {author} {\bibfnamefont {M.}~\bibnamefont {Rigobello}}, \bibinfo {author} {\bibfnamefont {S.}~\bibnamefont {Notarnicola}}, \bibinfo {author} {\bibfnamefont {G.}~\bibnamefont {Magnifico}},\ and\ \bibinfo {author} {\bibfnamefont {S.}~\bibnamefont {Montangero}},\ }\bibfield  {title} {\bibinfo {title} {Entanglement generation in 1+1d {QED} scattering processes},\ }\href {https://doi.org/10.1103/physrevd.104.114501} {\bibfield  {journal} {\bibinfo  {journal} {Phys. Rev. D}\ }\textbf {\bibinfo {volume} {104}},\ \bibinfo {pages} {114501} (\bibinfo {year} {2021})}\BibitemShut {NoStop}%
\bibitem [{\citenamefont {Rigobello}\ \emph {et~al.}(2023)\citenamefont {Rigobello}, \citenamefont {Magnifico}, \citenamefont {Silvi},\ and\ \citenamefont {Montangero}}]{Rigobello2023}%
  \BibitemOpen
  \bibfield  {author} {\bibinfo {author} {\bibfnamefont {M.}~\bibnamefont {Rigobello}}, \bibinfo {author} {\bibfnamefont {G.}~\bibnamefont {Magnifico}}, \bibinfo {author} {\bibfnamefont {P.}~\bibnamefont {Silvi}},\ and\ \bibinfo {author} {\bibfnamefont {S.}~\bibnamefont {Montangero}},\ }\bibfield  {title} {\bibinfo {title} {Hadrons in (1+1)d hamiltonian hardcore lattice qcd},\ }\href {https://doi.org/10.48550/ARXIV.2308.04488} {\bibfield  {journal} {\bibinfo  {journal} {arXiv:2308.04488}\ ,\ \bibinfo {pages} {{}}} (\bibinfo {year} {2023})}\BibitemShut {NoStop}%
\bibitem [{\citenamefont {Papaefstathiou}\ \emph {et~al.}(2025)\citenamefont {Papaefstathiou}, \citenamefont {Knolle},\ and\ \citenamefont {Ba\~nuls}}]{Papaefstathiou2024}%
  \BibitemOpen
  \bibfield  {author} {\bibinfo {author} {\bibfnamefont {I.}~\bibnamefont {Papaefstathiou}}, \bibinfo {author} {\bibfnamefont {J.}~\bibnamefont {Knolle}},\ and\ \bibinfo {author} {\bibfnamefont {M.~C.}\ \bibnamefont {Ba\~nuls}},\ }\bibfield  {title} {\bibinfo {title} {Real-time scattering in the lattice schwinger model},\ }\href {https://doi.org/10.1103/PhysRevD.111.014504} {\bibfield  {journal} {\bibinfo  {journal} {Phys. Rev. D}\ }\textbf {\bibinfo {volume} {111}},\ \bibinfo {pages} {014504} (\bibinfo {year} {2025})}\BibitemShut {NoStop}%
\bibitem [{\citenamefont {Angelides}\ \emph {et~al.}(2025{\natexlab{a}})\citenamefont {Angelides}, \citenamefont {Guo}, \citenamefont {Jansen}, \citenamefont {Kühn},\ and\ \citenamefont {Magnifico}}]{Angelides2025a}%
  \BibitemOpen
  \bibfield  {author} {\bibinfo {author} {\bibfnamefont {T.}~\bibnamefont {Angelides}}, \bibinfo {author} {\bibfnamefont {Y.}~\bibnamefont {Guo}}, \bibinfo {author} {\bibfnamefont {K.}~\bibnamefont {Jansen}}, \bibinfo {author} {\bibfnamefont {S.}~\bibnamefont {Kühn}},\ and\ \bibinfo {author} {\bibfnamefont {G.}~\bibnamefont {Magnifico}},\ }\bibfield  {title} {\bibinfo {title} {Meson thermalization with a hot medium in the open schwinger model},\ }\href {https://doi.org/10.1007/jhep04(2025)195} {\bibfield  {journal} {\bibinfo  {journal} {J. High Energy Phys.}\ }\textbf {\bibinfo {volume} {2025}}\bibinfo  {number} { (4)},\ \bibinfo {pages} {195}}\BibitemShut {NoStop}%
\bibitem [{\citenamefont {Chai}\ \emph {et~al.}(2025)\citenamefont {Chai}, \citenamefont {Guo},\ and\ \citenamefont {Kühn}}]{Chai2025}%
  \BibitemOpen
\bibfield  {number} {  }\bibfield  {author} {\bibinfo {author} {\bibfnamefont {Y.}~\bibnamefont {Chai}}, \bibinfo {author} {\bibfnamefont {Y.}~\bibnamefont {Guo}},\ and\ \bibinfo {author} {\bibfnamefont {S.}~\bibnamefont {Kühn}},\ }\bibfield  {title} {\bibinfo {title} {Scalable quantum algorithm for meson scattering in a lattice gauge theory},\ }\href {https://doi.org/10.48550/ARXIV.2505.21240} {\bibfield  {journal} {\bibinfo  {journal} {arXiv:2505.21240}\ ,\ \bibinfo {pages} {{}}} (\bibinfo {year} {2025})}\BibitemShut {NoStop}%
\bibitem [{\citenamefont {Felser}\ \emph {et~al.}(2020)\citenamefont {Felser}, \citenamefont {Silvi}, \citenamefont {Collura},\ and\ \citenamefont {Montangero}}]{Felser2019}%
  \BibitemOpen
  \bibfield  {author} {\bibinfo {author} {\bibfnamefont {T.}~\bibnamefont {Felser}}, \bibinfo {author} {\bibfnamefont {P.}~\bibnamefont {Silvi}}, \bibinfo {author} {\bibfnamefont {M.}~\bibnamefont {Collura}},\ and\ \bibinfo {author} {\bibfnamefont {S.}~\bibnamefont {Montangero}},\ }\bibfield  {title} {\bibinfo {title} {{Two-Dimensional Quantum-Link Lattice Quantum Electrodynamics at Finite Density}},\ }\href {https://doi.org/10.1103/PhysRevX.10.041040} {\bibfield  {journal} {\bibinfo  {journal} {Phys. Rev. X}\ }\textbf {\bibinfo {volume} {10}},\ \bibinfo {pages} {041040} (\bibinfo {year} {2020})}\BibitemShut {NoStop}%
\bibitem [{\citenamefont {Magnifico}\ \emph {et~al.}(2021)\citenamefont {Magnifico}, \citenamefont {Felser}, \citenamefont {Silvi},\ and\ \citenamefont {Montangero}}]{Magnifico2021}%
  \BibitemOpen
  \bibfield  {author} {\bibinfo {author} {\bibfnamefont {G.}~\bibnamefont {Magnifico}}, \bibinfo {author} {\bibfnamefont {T.}~\bibnamefont {Felser}}, \bibinfo {author} {\bibfnamefont {P.}~\bibnamefont {Silvi}},\ and\ \bibinfo {author} {\bibfnamefont {S.}~\bibnamefont {Montangero}},\ }\bibfield  {title} {{\selectlanguage {english}\bibinfo {title} {Lattice quantum electrodynamics in (3+1)-dimensions at finite density with tensor networks}},\ }\href {https://doi.org/10.1038/s41467-021-23646-3} {\bibfield  {journal} {\bibinfo  {journal} {Nat. Commun.}\ }\textbf {\bibinfo {volume} {12}},\ \bibinfo {pages} {3600} (\bibinfo {year} {2021})}\BibitemShut {NoStop}%
\bibitem [{\citenamefont {Cataldi}\ \emph {et~al.}(2024)\citenamefont {Cataldi}, \citenamefont {Magnifico}, \citenamefont {Silvi},\ and\ \citenamefont {Montangero}}]{Cataldi2024}%
  \BibitemOpen
  \bibfield  {author} {\bibinfo {author} {\bibfnamefont {G.}~\bibnamefont {Cataldi}}, \bibinfo {author} {\bibfnamefont {G.}~\bibnamefont {Magnifico}}, \bibinfo {author} {\bibfnamefont {P.}~\bibnamefont {Silvi}},\ and\ \bibinfo {author} {\bibfnamefont {S.}~\bibnamefont {Montangero}},\ }\bibfield  {title} {\bibinfo {title} {Simulating (2+1)d su(2) yang-mills lattice gauge theory at finite density with tensor networks},\ }\href {https://doi.org/10.1103/physrevresearch.6.033057} {\bibfield  {journal} {\bibinfo  {journal} {Physical Review Research}\ }\textbf {\bibinfo {volume} {6}},\ \bibinfo {pages} {033057} (\bibinfo {year} {2024})}\BibitemShut {NoStop}%
\bibitem [{\citenamefont {Magnifico}\ \emph {et~al.}(2025)\citenamefont {Magnifico}, \citenamefont {Cataldi}, \citenamefont {Rigobello}, \citenamefont {Majcen}, \citenamefont {Jaschke}, \citenamefont {Silvi},\ and\ \citenamefont {Montangero}}]{Magnifico2024}%
  \BibitemOpen
  \bibfield  {author} {\bibinfo {author} {\bibfnamefont {G.}~\bibnamefont {Magnifico}}, \bibinfo {author} {\bibfnamefont {G.}~\bibnamefont {Cataldi}}, \bibinfo {author} {\bibfnamefont {M.}~\bibnamefont {Rigobello}}, \bibinfo {author} {\bibfnamefont {P.}~\bibnamefont {Majcen}}, \bibinfo {author} {\bibfnamefont {D.}~\bibnamefont {Jaschke}}, \bibinfo {author} {\bibfnamefont {P.}~\bibnamefont {Silvi}},\ and\ \bibinfo {author} {\bibfnamefont {S.}~\bibnamefont {Montangero}},\ }\href {https://doi.org/10.1038/s42005-025-02125-x} {\bibinfo {title} {Tensor networks for lattice gauge theories beyond one dimension}} (\bibinfo {year} {2025})\BibitemShut {NoStop}%
\bibitem [{\citenamefont {Ginsparg}\ and\ \citenamefont {Wilson}(1982)}]{Ginsparg1982}%
  \BibitemOpen
  \bibfield  {author} {\bibinfo {author} {\bibfnamefont {P.~H.}\ \bibnamefont {Ginsparg}}\ and\ \bibinfo {author} {\bibfnamefont {K.~G.}\ \bibnamefont {Wilson}},\ }\bibfield  {title} {\bibinfo {title} {A remnant of chiral symmetry on the lattice},\ }\href {https://doi.org/10.1103/PhysRevD.25.2649} {\bibfield  {journal} {\bibinfo  {journal} {Phys. Rev. D}\ }\textbf {\bibinfo {volume} {25}},\ \bibinfo {pages} {2649} (\bibinfo {year} {1982})}\BibitemShut {NoStop}%
\bibitem [{\citenamefont {Aoki}\ and\ \citenamefont {B\"ar}(2006{\natexlab{a}})}]{Aoki1984}%
  \BibitemOpen
  \bibfield  {author} {\bibinfo {author} {\bibfnamefont {S.}~\bibnamefont {Aoki}}\ and\ \bibinfo {author} {\bibfnamefont {O.}~\bibnamefont {B\"ar}},\ }\bibfield  {title} {\bibinfo {title} {Automatic $o(a)$ improvement for twisted mass qcd in the presence of spontaneous symmetry breaking},\ }\href {https://doi.org/10.1103/PhysRevD.74.034511} {\bibfield  {journal} {\bibinfo  {journal} {Phys. Rev. D}\ }\textbf {\bibinfo {volume} {74}},\ \bibinfo {pages} {034511} (\bibinfo {year} {2006}{\natexlab{a}})}\BibitemShut {NoStop}%
\bibitem [{\citenamefont {L\"{u}scher}(1998)}]{Luescher1998}%
  \BibitemOpen
  \bibfield  {author} {\bibinfo {author} {\bibfnamefont {M.}~\bibnamefont {L\"{u}scher}},\ }\bibfield  {title} {\bibinfo {title} {Exact chiral symmetry on the lattice and the ginsparg-wilson relation},\ }\href {https://doi.org/10.1016/S0370-2693(98)00423-7} {\bibfield  {journal} {\bibinfo  {journal} {Phys. Lett. B}\ }\textbf {\bibinfo {volume} {428}},\ \bibinfo {pages} {342 } (\bibinfo {year} {1998})}\BibitemShut {NoStop}%
\bibitem [{\citenamefont {Frezzotti}\ \emph {et~al.}(2001)\citenamefont {Frezzotti}, \citenamefont {Grassi}, \citenamefont {Sint},\ and\ \citenamefont {Weisz}}]{Frezzotti2001}%
  \BibitemOpen
  \bibfield  {author} {\bibinfo {author} {\bibfnamefont {R.}~\bibnamefont {Frezzotti}}, \bibinfo {author} {\bibfnamefont {P.~A.}\ \bibnamefont {Grassi}}, \bibinfo {author} {\bibfnamefont {S.}~\bibnamefont {Sint}},\ and\ \bibinfo {author} {\bibfnamefont {P.}~\bibnamefont {Weisz}},\ }\bibfield  {title} {\bibinfo {title} {Lattice qcd with a chirally twisted mass term},\ }\href {http://stacks.iop.org/1126-6708/2001/i=08/a=058} {\bibfield  {journal} {\bibinfo  {journal} {J. High Energy Phys.}\ }\textbf {\bibinfo {volume} {2001}}\bibinfo  {number} { (08)},\ \bibinfo {pages} {058}}\BibitemShut {NoStop}%
\bibitem [{\citenamefont {Karsten}(1981)}]{Karsten1981}%
  \BibitemOpen
\bibfield  {number} {  }\bibfield  {author} {\bibinfo {author} {\bibfnamefont {L.~H.}\ \bibnamefont {Karsten}},\ }\bibfield  {title} {\bibinfo {title} {Lattice fermions in euclidean space-time},\ }\href {https://doi.org/http://dx.doi.org/10.1016/0370-2693(81)90133-7} {\bibfield  {journal} {\bibinfo  {journal} {Phys. Lett. B}\ }\textbf {\bibinfo {volume} {104}},\ \bibinfo {pages} {315 } (\bibinfo {year} {1981})}\BibitemShut {NoStop}%
\bibitem [{\citenamefont {Wilczek}(1987)}]{Wilczek1987}%
  \BibitemOpen
  \bibfield  {author} {\bibinfo {author} {\bibfnamefont {F.}~\bibnamefont {Wilczek}},\ }\bibfield  {title} {\bibinfo {title} {Lattice fermions},\ }\href {https://doi.org/10.1103/PhysRevLett.59.2397} {\bibfield  {journal} {\bibinfo  {journal} {Phys. Rev. Lett.}\ }\textbf {\bibinfo {volume} {59}},\ \bibinfo {pages} {2397} (\bibinfo {year} {1987})}\BibitemShut {NoStop}%
\bibitem [{\citenamefont {Bori\ifmmode~\mbox{\c{c}}\else\c{c}\fi{}i}(2008)}]{Borici2008}%
  \BibitemOpen
  \bibfield  {author} {\bibinfo {author} {\bibfnamefont {A.}~\bibnamefont {Bori\ifmmode~\mbox{\c{c}}\else\c{c}\fi{}i}},\ }\bibfield  {title} {\bibinfo {title} {Creutz fermions on an orthogonal lattice},\ }\href {https://doi.org/10.1103/PhysRevD.78.074504} {\bibfield  {journal} {\bibinfo  {journal} {Phys. Rev. D}\ }\textbf {\bibinfo {volume} {78}},\ \bibinfo {pages} {074504} (\bibinfo {year} {2008})}\BibitemShut {NoStop}%
\bibitem [{\citenamefont {Quinn}\ and\ \citenamefont {Weinstein}(1986)}]{Quinn1986}%
  \BibitemOpen
  \bibfield  {author} {\bibinfo {author} {\bibfnamefont {H.~R.}\ \bibnamefont {Quinn}}\ and\ \bibinfo {author} {\bibfnamefont {M.}~\bibnamefont {Weinstein}},\ }\bibfield  {title} {\bibinfo {title} {New formulation for the lattice-fermion derivative: Locality and chirality without spectrum doubling},\ }\href {https://doi.org/10.1103/PhysRevLett.57.2617} {\bibfield  {journal} {\bibinfo  {journal} {Phys. Rev. Lett.}\ }\textbf {\bibinfo {volume} {57}},\ \bibinfo {pages} {2617} (\bibinfo {year} {1986})}\BibitemShut {NoStop}%
\bibitem [{\citenamefont {Weber}(2015)}]{Weber2015}%
  \BibitemOpen
  \bibfield  {author} {\bibinfo {author} {\bibfnamefont {J.~H.}\ \bibnamefont {Weber}},\ }\emph {\bibinfo {title} {Properties of minimally doubled fermions}},\ \href {https://arxiv.org/abs/1706.07104} {Ph.D. thesis},\ \bibinfo  {school} {Johannes Gutenberg-Universit{\"a}t in Mainz} (\bibinfo {year} {2015})\BibitemShut {NoStop}%
\bibitem [{\citenamefont {Angelides}\ \emph {et~al.}(2023)\citenamefont {Angelides}, \citenamefont {Funcke}, \citenamefont {Jansen},\ and\ \citenamefont {Kühn}}]{Angelides2023}%
  \BibitemOpen
  \bibfield  {author} {\bibinfo {author} {\bibfnamefont {T.}~\bibnamefont {Angelides}}, \bibinfo {author} {\bibfnamefont {L.}~\bibnamefont {Funcke}}, \bibinfo {author} {\bibfnamefont {K.}~\bibnamefont {Jansen}},\ and\ \bibinfo {author} {\bibfnamefont {S.}~\bibnamefont {Kühn}},\ }\bibfield  {title} {\bibinfo {title} {Computing the {Mass} {Shift} of {Wilson} and {Staggered} {Fermions} in the {Lattice} {Schwinger} {Model} with {Matrix} {Product} {States}},\ }\bibfield  {journal} {\bibinfo  {journal} {Phys. Rev. D}\ }\textbf {\bibinfo {volume} {108}},\ \href {https://doi.org/10.1103/PhysRevD.108.014516} {10.1103/PhysRevD.108.014516} (\bibinfo {year} {2023}),\ \bibinfo {note} {arXiv:2303.11016 [hep-lat]}\BibitemShut {NoStop}%
\bibitem [{\citenamefont {Shindler}(2008)}]{Shindler2008}%
  \BibitemOpen
  \bibfield  {author} {\bibinfo {author} {\bibfnamefont {A.}~\bibnamefont {Shindler}},\ }\bibfield  {title} {\bibinfo {title} {Twisted mass lattice {QCD}},\ }\href {https://doi.org/10.1016/j.physrep.2008.03.001} {\bibfield  {journal} {\bibinfo  {journal} {Physics Reports}\ }\textbf {\bibinfo {volume} {461}},\ \bibinfo {pages} {37} (\bibinfo {year} {2008})},\ \bibinfo {note} {arXiv:0707.4093 [hep-lat]}\BibitemShut {NoStop}%
\bibitem [{\citenamefont {Albandea}\ and\ \citenamefont {Hernández}(2025)}]{Albandea2025}%
  \BibitemOpen
  \bibfield  {author} {\bibinfo {author} {\bibfnamefont {D.}~\bibnamefont {Albandea}}\ and\ \bibinfo {author} {\bibfnamefont {P.}~\bibnamefont {Hernández}},\ }\bibfield  {title} {\bibinfo {title} {Chiral and isospin breaking in the two-flavor {Schwinger} model},\ }\href {https://doi.org/10.1103/PhysRevD.111.074503} {\bibfield  {journal} {\bibinfo  {journal} {Phys. Rev. D}\ }\textbf {\bibinfo {volume} {111}},\ \bibinfo {pages} {074503} (\bibinfo {year} {2025})}\BibitemShut {NoStop}%
\bibitem [{\citenamefont {Wilson}(1974)}]{Wilson1974}%
  \BibitemOpen
  \bibfield  {author} {\bibinfo {author} {\bibfnamefont {K.~G.}\ \bibnamefont {Wilson}},\ }\bibfield  {title} {\bibinfo {title} {Confinement of quarks},\ }\href {https://doi.org/10.1103/PhysRevD.10.2445} {\bibfield  {journal} {\bibinfo  {journal} {Phys. Rev. D}\ }\textbf {\bibinfo {volume} {10}},\ \bibinfo {pages} {2445} (\bibinfo {year} {1974})}\BibitemShut {NoStop}%
\bibitem [{\citenamefont {Bañuls}\ \emph {et~al.}(2016)\citenamefont {Bañuls}, \citenamefont {Cichy}, \citenamefont {Jansen},\ and\ \citenamefont {Saito}}]{Banuls2016}%
  \BibitemOpen
  \bibfield  {author} {\bibinfo {author} {\bibfnamefont {M.~C.}\ \bibnamefont {Bañuls}}, \bibinfo {author} {\bibfnamefont {K.}~\bibnamefont {Cichy}}, \bibinfo {author} {\bibfnamefont {K.}~\bibnamefont {Jansen}},\ and\ \bibinfo {author} {\bibfnamefont {H.}~\bibnamefont {Saito}},\ }\bibfield  {title} {\bibinfo {title} {Chiral condensate in the {Schwinger} model with {Matrix} {Product} {Operators}},\ }\bibfield  {journal} {\bibinfo  {journal} {Phys. Rev. D}\ }\textbf {\bibinfo {volume} {93}},\ \href {https://doi.org/10.1103/PhysRevD.93.094512} {10.1103/PhysRevD.93.094512} (\bibinfo {year} {2016}),\ \bibinfo {note} {arXiv:1603.05002 [hep-lat]}\BibitemShut {NoStop}%
\bibitem [{\citenamefont {Funcke}\ \emph {et~al.}(2020)\citenamefont {Funcke}, \citenamefont {Jansen},\ and\ \citenamefont {Kühn}}]{Funcke2020}%
  \BibitemOpen
  \bibfield  {author} {\bibinfo {author} {\bibfnamefont {L.}~\bibnamefont {Funcke}}, \bibinfo {author} {\bibfnamefont {K.}~\bibnamefont {Jansen}},\ and\ \bibinfo {author} {\bibfnamefont {S.}~\bibnamefont {Kühn}},\ }\bibfield  {title} {\bibinfo {title} {Topological vacuum structure of the {Schwinger} model with matrix product states},\ }\href {https://doi.org/10.1103/PhysRevD.101.054507} {\bibfield  {journal} {\bibinfo  {journal} {Phys. Rev. D}\ }\textbf {\bibinfo {volume} {101}},\ \bibinfo {pages} {054507} (\bibinfo {year} {2020})},\ \bibinfo {note} {arXiv:1908.00551 [hep-lat]}\BibitemShut {NoStop}%
\bibitem [{\citenamefont {Zache}\ \emph {et~al.}(2022)\citenamefont {Zache}, \citenamefont {Van~Damme}, \citenamefont {Halimeh}, \citenamefont {Hauke},\ and\ \citenamefont {Banerjee}}]{Zache2022}%
  \BibitemOpen
  \bibfield  {author} {\bibinfo {author} {\bibfnamefont {T.~V.}\ \bibnamefont {Zache}}, \bibinfo {author} {\bibfnamefont {M.}~\bibnamefont {Van~Damme}}, \bibinfo {author} {\bibfnamefont {J.~C.}\ \bibnamefont {Halimeh}}, \bibinfo {author} {\bibfnamefont {P.}~\bibnamefont {Hauke}},\ and\ \bibinfo {author} {\bibfnamefont {D.}~\bibnamefont {Banerjee}},\ }\bibfield  {title} {\bibinfo {title} {Toward the continuum limit of a (1+1)\text{D} quantum link schwinger model},\ }\href {https://doi.org/10.1103/PhysRevD.106.L091502} {\bibfield  {journal} {\bibinfo  {journal} {Phys. Rev. D}\ }\textbf {\bibinfo {volume} {106}},\ \bibinfo {pages} {L091502} (\bibinfo {year} {2022})}\BibitemShut {NoStop}%
\bibitem [{\citenamefont {Bañuls}\ \emph {et~al.}(2017)\citenamefont {Bañuls}, \citenamefont {Cichy}, \citenamefont {Cirac}, \citenamefont {Jansen},\ and\ \citenamefont {Kühn}}]{Banuls2017}%
  \BibitemOpen
  \bibfield  {author} {\bibinfo {author} {\bibfnamefont {M.~C.}\ \bibnamefont {Bañuls}}, \bibinfo {author} {\bibfnamefont {K.}~\bibnamefont {Cichy}}, \bibinfo {author} {\bibfnamefont {J.~I.}\ \bibnamefont {Cirac}}, \bibinfo {author} {\bibfnamefont {K.}~\bibnamefont {Jansen}},\ and\ \bibinfo {author} {\bibfnamefont {S.}~\bibnamefont {Kühn}},\ }\bibfield  {title} {\bibinfo {title} {Density {Induced} {Phase} {Transitions} in the {Schwinger} {Model}: {A} {Study} with {Matrix} {Product} {States}},\ }\href {https://doi.org/10.1103/PhysRevLett.118.071601} {\bibfield  {journal} {\bibinfo  {journal} {Phys. Rev. Lett.}\ }\textbf {\bibinfo {volume} {118}},\ \bibinfo {pages} {071601} (\bibinfo {year} {2017})},\ \bibinfo {note} {arXiv:1611.00705 [hep-lat]}\BibitemShut {NoStop}%
\bibitem [{\citenamefont {Itou}\ \emph {et~al.}(2023)\citenamefont {Itou}, \citenamefont {Matsumoto},\ and\ \citenamefont {Tanizaki}}]{Itou2023}%
  \BibitemOpen
  \bibfield  {author} {\bibinfo {author} {\bibfnamefont {E.}~\bibnamefont {Itou}}, \bibinfo {author} {\bibfnamefont {A.}~\bibnamefont {Matsumoto}},\ and\ \bibinfo {author} {\bibfnamefont {Y.}~\bibnamefont {Tanizaki}},\ }\href {https://doi.org/10.48550/arXiv.2307.16655} {\bibinfo {title} {Calculating composite-particle spectra in {Hamiltonian} formalism and demonstration in 2-flavor {QED}$_{1+1d}$}} (\bibinfo {year} {2023}),\ \bibinfo {note} {arXiv:2307.16655 [hep-lat]}\BibitemShut {NoStop}%
\bibitem [{\citenamefont {Funcke}\ \emph {et~al.}(2023)\citenamefont {Funcke}, \citenamefont {Jansen},\ and\ \citenamefont {Kühn}}]{Funcke2023}%
  \BibitemOpen
  \bibfield  {author} {\bibinfo {author} {\bibfnamefont {L.}~\bibnamefont {Funcke}}, \bibinfo {author} {\bibfnamefont {K.}~\bibnamefont {Jansen}},\ and\ \bibinfo {author} {\bibfnamefont {S.}~\bibnamefont {Kühn}},\ }\bibfield  {title} {\bibinfo {title} {Exploring the {CP}-violating {Dashen} phase in the {Schwinger} model with tensor networks},\ }\href {https://doi.org/10.1103/PhysRevD.108.014504} {\bibfield  {journal} {\bibinfo  {journal} {Phys. Rev. D}\ }\textbf {\bibinfo {volume} {108}},\ \bibinfo {pages} {014504} (\bibinfo {year} {2023})},\ \bibinfo {note} {arXiv:2303.03799 [hep-lat]}\BibitemShut {NoStop}%
\bibitem [{\citenamefont {Itou}\ \emph {et~al.}(2024)\citenamefont {Itou}, \citenamefont {Matsumoto},\ and\ \citenamefont {Tanizaki}}]{Itou2024}%
  \BibitemOpen
  \bibfield  {author} {\bibinfo {author} {\bibfnamefont {E.}~\bibnamefont {Itou}}, \bibinfo {author} {\bibfnamefont {A.}~\bibnamefont {Matsumoto}},\ and\ \bibinfo {author} {\bibfnamefont {Y.}~\bibnamefont {Tanizaki}},\ }\href {https://doi.org/10.48550/arXiv.2407.11391} {\bibinfo {title} {{DMRG} study of the theta-dependent mass spectrum in the 2-flavor {Schwinger} model}} (\bibinfo {year} {2024}),\ \bibinfo {note} {arXiv:2407.11391 [hep-lat]}\BibitemShut {NoStop}%
\bibitem [{\citenamefont {Dempsey}\ \emph {et~al.}(2024)\citenamefont {Dempsey}, \citenamefont {Klebanov}, \citenamefont {Pufu}, \citenamefont {Søgaard},\ and\ \citenamefont {Zan}}]{Dempsey2024}%
  \BibitemOpen
  \bibfield  {author} {\bibinfo {author} {\bibfnamefont {R.}~\bibnamefont {Dempsey}}, \bibinfo {author} {\bibfnamefont {I.~R.}\ \bibnamefont {Klebanov}}, \bibinfo {author} {\bibfnamefont {S.~S.}\ \bibnamefont {Pufu}}, \bibinfo {author} {\bibfnamefont {B.~T.}\ \bibnamefont {Søgaard}},\ and\ \bibinfo {author} {\bibfnamefont {B.}~\bibnamefont {Zan}},\ }\bibfield  {title} {\bibinfo {title} {Phase {Diagram} of the {Two}-{Flavor} {Schwinger} {Model} at {Zero} {Temperature}},\ }\href {https://doi.org/10.1103/PhysRevLett.132.031603} {\bibfield  {journal} {\bibinfo  {journal} {Phys. Rev. Lett.}\ }\textbf {\bibinfo {volume} {132}},\ \bibinfo {pages} {031603} (\bibinfo {year} {2024})},\ \bibinfo {note} {arXiv:2305.04437 [hep-th]}\BibitemShut {NoStop}%
\bibitem [{\citenamefont {Kühn}\ \emph {et~al.}(2022)\citenamefont {Kühn}, \citenamefont {Funcke}, \citenamefont {Hartung}, \citenamefont {Jansen}, \citenamefont {Pleinert}, \citenamefont {Schuster},\ and\ \citenamefont {von Zanthier}}]{Kuehn2022}%
  \BibitemOpen
  \bibfield  {author} {\bibinfo {author} {\bibfnamefont {S.}~\bibnamefont {Kühn}}, \bibinfo {author} {\bibfnamefont {L.}~\bibnamefont {Funcke}}, \bibinfo {author} {\bibfnamefont {T.}~\bibnamefont {Hartung}}, \bibinfo {author} {\bibfnamefont {K.}~\bibnamefont {Jansen}}, \bibinfo {author} {\bibfnamefont {M.-O.}\ \bibnamefont {Pleinert}}, \bibinfo {author} {\bibfnamefont {S.}~\bibnamefont {Schuster}},\ and\ \bibinfo {author} {\bibfnamefont {J.}~\bibnamefont {von Zanthier}},\ }\href {https://doi.org/10.48550/arXiv.2211.13020} {\bibinfo {title} {Exploring the phase structure of the multi-flavor {Schwinger} model with quantum computing}} (\bibinfo {year} {2022}),\ \bibinfo {note} {arXiv:2211.13020 [quant-ph]}\BibitemShut {NoStop}%
\bibitem [{\citenamefont {Farrell}\ \emph {et~al.}(2024)\citenamefont {Farrell}, \citenamefont {Illa}, \citenamefont {Ciavarella},\ and\ \citenamefont {Savage}}]{Farrell2024}%
  \BibitemOpen
  \bibfield  {author} {\bibinfo {author} {\bibfnamefont {R.~C.}\ \bibnamefont {Farrell}}, \bibinfo {author} {\bibfnamefont {M.}~\bibnamefont {Illa}}, \bibinfo {author} {\bibfnamefont {A.~N.}\ \bibnamefont {Ciavarella}},\ and\ \bibinfo {author} {\bibfnamefont {M.~J.}\ \bibnamefont {Savage}},\ }\bibfield  {title} {\bibinfo {title} {Scalable {Circuits} for {Preparing} {Ground} {States} on {Digital} {Quantum} {Computers}: {The} {Schwinger} {Model} {Vacuum} on 100 {Qubits}},\ }\href {https://doi.org/10.1103/PRXQuantum.5.020315} {\bibfield  {journal} {\bibinfo  {journal} {PRX Quantum}\ }\textbf {\bibinfo {volume} {5}},\ \bibinfo {pages} {020315} (\bibinfo {year} {2024})}\BibitemShut {NoStop}%
\bibitem [{\citenamefont {Bringewatt}\ \emph {et~al.}(2024)\citenamefont {Bringewatt}, \citenamefont {Kunjummen},\ and\ \citenamefont {Mueller}}]{Bringewatt2024}%
  \BibitemOpen
  \bibfield  {author} {\bibinfo {author} {\bibfnamefont {J.}~\bibnamefont {Bringewatt}}, \bibinfo {author} {\bibfnamefont {J.}~\bibnamefont {Kunjummen}},\ and\ \bibinfo {author} {\bibfnamefont {N.}~\bibnamefont {Mueller}},\ }\href {https://doi.org/10.48550/arXiv.2403.08859} {\bibinfo {title} {Solving lattice gauge theories using the quantum {Krylov} algorithm and qubitization}} (\bibinfo {year} {2024}),\ \bibinfo {note} {arXiv:2403.08859 [quant-ph]}\BibitemShut {NoStop}%
\bibitem [{\citenamefont {Angelides}\ \emph {et~al.}(2025{\natexlab{b}})\citenamefont {Angelides}, \citenamefont {Naredi}, \citenamefont {Crippa}, \citenamefont {Jansen}, \citenamefont {Kühn}, \citenamefont {Tavernelli},\ and\ \citenamefont {Wang}}]{Angelides2025}%
  \BibitemOpen
  \bibfield  {author} {\bibinfo {author} {\bibfnamefont {T.}~\bibnamefont {Angelides}}, \bibinfo {author} {\bibfnamefont {P.}~\bibnamefont {Naredi}}, \bibinfo {author} {\bibfnamefont {A.}~\bibnamefont {Crippa}}, \bibinfo {author} {\bibfnamefont {K.}~\bibnamefont {Jansen}}, \bibinfo {author} {\bibfnamefont {S.}~\bibnamefont {Kühn}}, \bibinfo {author} {\bibfnamefont {I.}~\bibnamefont {Tavernelli}},\ and\ \bibinfo {author} {\bibfnamefont {D.~S.}\ \bibnamefont {Wang}},\ }\bibfield  {title} {{\selectlanguage {english}\bibinfo {title} {First-order phase transition of the {Schwinger} model with a quantum computer}},\ }\href {https://doi.org/10.1038/s41534-024-00950-6} {\bibfield  {journal} {\bibinfo  {journal} {npj Quantum Inf.}\ }\textbf {\bibinfo {volume} {11}},\ \bibinfo {pages} {1} (\bibinfo {year} {2025}{\natexlab{b}})}\BibitemShut {NoStop}%
\bibitem [{\citenamefont {Schuster}\ \emph {et~al.}(2024)\citenamefont {Schuster}, \citenamefont {Kühn}, \citenamefont {Funcke}, \citenamefont {Hartung}, \citenamefont {Pleinert}, \citenamefont {Zanthier},\ and\ \citenamefont {Jansen}}]{Schuster2024}%
  \BibitemOpen
  \bibfield  {author} {\bibinfo {author} {\bibfnamefont {S.}~\bibnamefont {Schuster}}, \bibinfo {author} {\bibfnamefont {S.}~\bibnamefont {Kühn}}, \bibinfo {author} {\bibfnamefont {L.}~\bibnamefont {Funcke}}, \bibinfo {author} {\bibfnamefont {T.}~\bibnamefont {Hartung}}, \bibinfo {author} {\bibfnamefont {M.-O.}\ \bibnamefont {Pleinert}}, \bibinfo {author} {\bibfnamefont {J.~v.}\ \bibnamefont {Zanthier}},\ and\ \bibinfo {author} {\bibfnamefont {K.}~\bibnamefont {Jansen}},\ }\bibfield  {title} {\bibinfo {title} {Studying the phase diagram of the three-flavor {Schwinger} model in the presence of a chemical potential with measurement- and gate-based quantum computing},\ }\href {https://doi.org/10.1103/PhysRevD.109.114508} {\bibfield  {journal} {\bibinfo  {journal} {Phys. Rev. D}\ }\textbf {\bibinfo {volume} {109}},\ \bibinfo {pages} {114508} (\bibinfo {year} {2024})},\ \bibinfo {note} {arXiv:2311.14825 [hep-lat]}\BibitemShut {NoStop}%
\bibitem [{\citenamefont {Guo}\ \emph {et~al.}(2024)\citenamefont {Guo}, \citenamefont {Angelides}, \citenamefont {Jansen},\ and\ \citenamefont {Kühn}}]{guo2024}%
  \BibitemOpen
  \bibfield  {author} {\bibinfo {author} {\bibfnamefont {Y.}~\bibnamefont {Guo}}, \bibinfo {author} {\bibfnamefont {T.}~\bibnamefont {Angelides}}, \bibinfo {author} {\bibfnamefont {K.}~\bibnamefont {Jansen}},\ and\ \bibinfo {author} {\bibfnamefont {S.}~\bibnamefont {Kühn}},\ }\href {https://arxiv.org/abs/2407.15629} {\bibinfo {title} {Concurrent vqe for simulating excited states of the schwinger model}} (\bibinfo {year} {2024}),\ \Eprint {https://arxiv.org/abs/2407.15629} {arXiv:2407.15629 [quant-ph]} \BibitemShut {NoStop}%
\bibitem [{\citenamefont {Christian}\ \emph {et~al.}(2006)\citenamefont {Christian}, \citenamefont {Jansen}, \citenamefont {Nagai},\ and\ \citenamefont {Pollakowski}}]{Christian2006}%
  \BibitemOpen
  \bibfield  {author} {\bibinfo {author} {\bibfnamefont {N.}~\bibnamefont {Christian}}, \bibinfo {author} {\bibfnamefont {K.}~\bibnamefont {Jansen}}, \bibinfo {author} {\bibfnamefont {K.}~\bibnamefont {Nagai}},\ and\ \bibinfo {author} {\bibfnamefont {B.}~\bibnamefont {Pollakowski}},\ }\bibfield  {title} {\bibinfo {title} {Scaling test of fermion actions in the {Schwinger} model},\ }\href {https://doi.org/10.1016/j.nuclphysb.2006.01.029} {\bibfield  {journal} {\bibinfo  {journal} {Nucl. Phys. B}\ }\textbf {\bibinfo {volume} {739}},\ \bibinfo {pages} {60} (\bibinfo {year} {2006})},\ \bibinfo {note} {arXiv:hep-lat/0510047}\BibitemShut {NoStop}%
\bibitem [{\citenamefont {Dempsey}\ \emph {et~al.}(2022)\citenamefont {Dempsey}, \citenamefont {Klebanov}, \citenamefont {Pufu},\ and\ \citenamefont {Zan}}]{Dempsey2022}%
  \BibitemOpen
  \bibfield  {author} {\bibinfo {author} {\bibfnamefont {R.}~\bibnamefont {Dempsey}}, \bibinfo {author} {\bibfnamefont {I.~R.}\ \bibnamefont {Klebanov}}, \bibinfo {author} {\bibfnamefont {S.~S.}\ \bibnamefont {Pufu}},\ and\ \bibinfo {author} {\bibfnamefont {B.}~\bibnamefont {Zan}},\ }\bibfield  {title} {\bibinfo {title} {Discrete chiral symmetry and mass shift in the lattice {Hamiltonian} approach to the {Schwinger} model},\ }\href {https://doi.org/10.1103/PhysRevResearch.4.043133} {\bibfield  {journal} {\bibinfo  {journal} {Physical Review Research}\ }\textbf {\bibinfo {volume} {4}},\ \bibinfo {pages} {043133} (\bibinfo {year} {2022})}\BibitemShut {NoStop}%
\bibitem [{\citenamefont {Bañuls}\ \emph {et~al.}(2013)\citenamefont {Bañuls}, \citenamefont {Cichy}, \citenamefont {Jansen},\ and\ \citenamefont {Cirac}}]{Banuls2013}%
  \BibitemOpen
  \bibfield  {author} {\bibinfo {author} {\bibfnamefont {M.~C.}\ \bibnamefont {Bañuls}}, \bibinfo {author} {\bibfnamefont {K.}~\bibnamefont {Cichy}}, \bibinfo {author} {\bibfnamefont {K.}~\bibnamefont {Jansen}},\ and\ \bibinfo {author} {\bibfnamefont {J.~I.}\ \bibnamefont {Cirac}},\ }\bibfield  {title} {\bibinfo {title} {The mass spectrum of the {Schwinger} model with {Matrix} {Product} {States}},\ }\href {https://doi.org/10.1007/JHEP11(2013)158} {\bibfield  {journal} {\bibinfo  {journal} {J. High Energy Phys.}\ }\textbf {\bibinfo {volume} {2013}}\bibfield  {number} {\bibinfo  {number} { (11)}},\ }\bibinfo {note} {arXiv:1305.3765 [hep-lat]}\BibitemShut {NoStop}%
\bibitem [{\citenamefont {Smilga}(1997)}]{Smilga1997}%
  \BibitemOpen
  \bibfield  {author} {\bibinfo {author} {\bibfnamefont {A.~V.}\ \bibnamefont {Smilga}},\ }\bibfield  {title} {\bibinfo {title} {Critical amplitudes in two-dimensional theories},\ }\href {https://doi.org/10.1103/PhysRevD.55.R443} {\bibfield  {journal} {\bibinfo  {journal} {Phys. Rev. D}\ }\textbf {\bibinfo {volume} {55}},\ \bibinfo {pages} {R443} (\bibinfo {year} {1997})}\BibitemShut {NoStop}%
\bibitem [{\citenamefont {Hosotani}\ and\ \citenamefont {Rodriguez}(1998)}]{Hosotani1998}%
  \BibitemOpen
  \bibfield  {author} {\bibinfo {author} {\bibfnamefont {Y.}~\bibnamefont {Hosotani}}\ and\ \bibinfo {author} {\bibfnamefont {R.}~\bibnamefont {Rodriguez}},\ }\bibfield  {title} {\bibinfo {title} {Bosonized {Massive} {N}-flavor {Schwinger} {Model}},\ }\href {https://doi.org/10.1088/0305-4470/31/49/013} {\bibfield  {journal} {\bibinfo  {journal} {Journal of Physics A: Mathematical and General}\ }\textbf {\bibinfo {volume} {31}},\ \bibinfo {pages} {9925} (\bibinfo {year} {1998})},\ \bibinfo {note} {arXiv:hep-th/9804205}\BibitemShut {NoStop}%
\bibitem [{\citenamefont {Schwinger}(1962)}]{Schwinger1962}%
  \BibitemOpen
  \bibfield  {author} {\bibinfo {author} {\bibfnamefont {J.}~\bibnamefont {Schwinger}},\ }\bibfield  {title} {\bibinfo {title} {Gauge {Invariance} and {Mass}. {II}},\ }\href {https://doi.org/10.1103/PhysRev.128.2425} {\bibfield  {journal} {\bibinfo  {journal} {Physical Review}\ }\textbf {\bibinfo {volume} {128}},\ \bibinfo {pages} {2425} (\bibinfo {year} {1962})}\BibitemShut {NoStop}%
\bibitem [{\citenamefont {Adam}(1997)}]{Adam1997}%
  \BibitemOpen
  \bibfield  {author} {\bibinfo {author} {\bibfnamefont {C.}~\bibnamefont {Adam}},\ }\bibfield  {title} {\bibinfo {title} {Massive {Schwinger} model within mass perturbation theory},\ }\href {https://doi.org/10.1006/aphy.1997.5697} {\bibfield  {journal} {\bibinfo  {journal} {Annals of Phys.}\ }\textbf {\bibinfo {volume} {259}},\ \bibinfo {pages} {1} (\bibinfo {year} {1997})},\ \bibinfo {note} {arXiv:hep-th/9704064}\BibitemShut {NoStop}%
\bibitem [{\citenamefont {Gepner}(1985)}]{Gepner1985}%
  \BibitemOpen
  \bibfield  {author} {\bibinfo {author} {\bibfnamefont {D.}~\bibnamefont {Gepner}},\ }\bibfield  {title} {\bibinfo {title} {Non-abelian bosonization and multiflavor {QED} and {QCD} in two dimensions},\ }\href {https://doi.org/10.1016/0550-3213(85)90458-4} {\bibfield  {journal} {\bibinfo  {journal} {Nucl. Phys. B}\ }\textbf {\bibinfo {volume} {252}},\ \bibinfo {pages} {481} (\bibinfo {year} {1985})}\BibitemShut {NoStop}%
\bibitem [{\citenamefont {Affleck}(1986)}]{Affleck1986}%
  \BibitemOpen
  \bibfield  {author} {\bibinfo {author} {\bibfnamefont {I.}~\bibnamefont {Affleck}},\ }\bibfield  {title} {\bibinfo {title} {On the realization of chiral symmetry in (1+1) dimensions},\ }\href {https://doi.org/10.1016/0550-3213(86)90168-9} {\bibfield  {journal} {\bibinfo  {journal} {Nucl. Phys. B}\ }\textbf {\bibinfo {volume} {265}},\ \bibinfo {pages} {448} (\bibinfo {year} {1986})}\BibitemShut {NoStop}%
\bibitem [{\citenamefont {Coleman}(1976)}]{Coleman1976}%
  \BibitemOpen
  \bibfield  {author} {\bibinfo {author} {\bibfnamefont {S.}~\bibnamefont {Coleman}},\ }\bibfield  {title} {\bibinfo {title} {More about the massive {Schwinger} model},\ }\href {https://doi.org/10.1016/0003-4916(76)90280-3} {\bibfield  {journal} {\bibinfo  {journal} {Annals of Phys.}\ }\textbf {\bibinfo {volume} {101}},\ \bibinfo {pages} {239} (\bibinfo {year} {1976})}\BibitemShut {NoStop}%
\bibitem [{\citenamefont {Hamer}\ \emph {et~al.}(1997)\citenamefont {Hamer}, \citenamefont {Weihong},\ and\ \citenamefont {Oitmaa}}]{Hamer1997a}%
  \BibitemOpen
  \bibfield  {author} {\bibinfo {author} {\bibfnamefont {C.~J.}\ \bibnamefont {Hamer}}, \bibinfo {author} {\bibfnamefont {Z.}~\bibnamefont {Weihong}},\ and\ \bibinfo {author} {\bibfnamefont {J.}~\bibnamefont {Oitmaa}},\ }\bibfield  {title} {\bibinfo {title} {Series {Expansions} for the {Massive} {Schwinger} {Model} in {Hamiltonian} lattice theory},\ }\href {https://doi.org/10.1103/PhysRevD.56.55} {\bibfield  {journal} {\bibinfo  {journal} {Phys. Rev. D}\ }\textbf {\bibinfo {volume} {56}},\ \bibinfo {pages} {55} (\bibinfo {year} {1997})},\ \bibinfo {note} {arXiv:hep-lat/9701015}\BibitemShut {NoStop}%
\bibitem [{\citenamefont {Zache}\ \emph {et~al.}(2018)\citenamefont {Zache}, \citenamefont {Hebenstreit}, \citenamefont {Jendrzejewski}, \citenamefont {Oberthaler}, \citenamefont {Berges},\ and\ \citenamefont {Hauke}}]{Zache2018}%
  \BibitemOpen
  \bibfield  {author} {\bibinfo {author} {\bibfnamefont {T.~V.}\ \bibnamefont {Zache}}, \bibinfo {author} {\bibfnamefont {F.}~\bibnamefont {Hebenstreit}}, \bibinfo {author} {\bibfnamefont {F.}~\bibnamefont {Jendrzejewski}}, \bibinfo {author} {\bibfnamefont {M.~K.}\ \bibnamefont {Oberthaler}}, \bibinfo {author} {\bibfnamefont {J.}~\bibnamefont {Berges}},\ and\ \bibinfo {author} {\bibfnamefont {P.}~\bibnamefont {Hauke}},\ }\bibfield  {title} {\bibinfo {title} {Quantum simulation of lattice gauge theories using {Wilson} fermions},\ }\href {https://doi.org/10.1088/2058-9565/aac33b} {\bibfield  {journal} {\bibinfo  {journal} {Quantum Science and Technology}\ }\textbf {\bibinfo {volume} {3}},\ \bibinfo {pages} {034010} (\bibinfo {year} {2018})},\ \bibinfo {note} {arXiv:1802.06704 [cond-mat]}\BibitemShut {NoStop}%
\bibitem [{\citenamefont {Mazzola}\ \emph {et~al.}(2021)\citenamefont {Mazzola}, \citenamefont {Mathis}, \citenamefont {Mazzola},\ and\ \citenamefont {Tavernelli}}]{Mazzola2021}%
  \BibitemOpen
  \bibfield  {author} {\bibinfo {author} {\bibfnamefont {G.}~\bibnamefont {Mazzola}}, \bibinfo {author} {\bibfnamefont {S.~V.}\ \bibnamefont {Mathis}}, \bibinfo {author} {\bibfnamefont {G.}~\bibnamefont {Mazzola}},\ and\ \bibinfo {author} {\bibfnamefont {I.}~\bibnamefont {Tavernelli}},\ }\bibfield  {title} {\bibinfo {title} {Gauge-invariant quantum circuits for u(1) and yang-mills lattice gauge theories},\ }\bibfield  {journal} {\bibinfo  {journal} {Physical Review Research}\ }\textbf {\bibinfo {volume} {3}},\ \href {https://doi.org/10.1103/physrevresearch.3.043209} {10.1103/physrevresearch.3.043209} (\bibinfo {year} {2021})\BibitemShut {NoStop}%
\bibitem [{\citenamefont {Frezzotti}\ and\ \citenamefont {Rossi}(2004)}]{Frezzotti2004}%
  \BibitemOpen
  \bibfield  {author} {\bibinfo {author} {\bibfnamefont {R.}~\bibnamefont {Frezzotti}}\ and\ \bibinfo {author} {\bibfnamefont {G.~C.}\ \bibnamefont {Rossi}},\ }\bibfield  {title} {\bibinfo {title} {Chirally improving {Wilson} fermions - {I}. {O}(a) improvement},\ }\href {https://doi.org/10.1088/1126-6708/2004/08/007} {\bibfield  {journal} {\bibinfo  {journal} {J. High Energy Phys.}\ }\textbf {\bibinfo {volume} {2004}}\bibfield  {number} {\bibinfo  {number} { (08)},\ \bibinfo {pages} {007}},\ }\bibinfo {note} {arXiv:hep-lat/0306014}\BibitemShut {NoStop}%
\bibitem [{\citenamefont {Aoki}\ and\ \citenamefont {Bär}(2004)}]{Aoki2004}%
  \BibitemOpen
  \bibfield  {author} {\bibinfo {author} {\bibfnamefont {S.}~\bibnamefont {Aoki}}\ and\ \bibinfo {author} {\bibfnamefont {O.}~\bibnamefont {Bär}},\ }\bibfield  {title} {\bibinfo {title} {Twisted mass {QCD}, {O}(a) improvement, and {Wilson} chiral perturbation theory},\ }\href {https://doi.org/10.1103/PhysRevD.70.116011} {\bibfield  {journal} {\bibinfo  {journal} {Phys. Rev. D}\ }\textbf {\bibinfo {volume} {70}},\ \bibinfo {pages} {116011} (\bibinfo {year} {2004})}\BibitemShut {NoStop}%
\bibitem [{\citenamefont {Aoki}\ and\ \citenamefont {B\"ar}(2006{\natexlab{b}})}]{Aoki2006}%
  \BibitemOpen
  \bibfield  {author} {\bibinfo {author} {\bibfnamefont {S.}~\bibnamefont {Aoki}}\ and\ \bibinfo {author} {\bibfnamefont {O.}~\bibnamefont {B\"ar}},\ }\bibfield  {title} {\bibinfo {title} {Automatic $o(a)$ improvement for twisted mass qcd in the presence of spontaneous symmetry breaking},\ }\href {https://doi.org/10.1103/PhysRevD.74.034511} {\bibfield  {journal} {\bibinfo  {journal} {Phys. Rev. D}\ }\textbf {\bibinfo {volume} {74}},\ \bibinfo {pages} {034511} (\bibinfo {year} {2006}{\natexlab{b}})}\BibitemShut {NoStop}%
\bibitem [{\citenamefont {Schollwöck}(2011)}]{Schollwoeck_2011}%
  \BibitemOpen
  \bibfield  {author} {\bibinfo {author} {\bibfnamefont {U.}~\bibnamefont {Schollwöck}},\ }\bibfield  {title} {\bibinfo {title} {The density-matrix renormalization group in the age of matrix product states},\ }\href {https://doi.org/10.1016/j.aop.2010.09.012} {\bibfield  {journal} {\bibinfo  {journal} {Annals of Physics}\ }\textbf {\bibinfo {volume} {326}},\ \bibinfo {pages} {96–192} (\bibinfo {year} {2011})}\BibitemShut {NoStop}%
\bibitem [{\citenamefont {Fishman}\ \emph {et~al.}(2022)\citenamefont {Fishman}, \citenamefont {White},\ and\ \citenamefont {Stoudenmire}}]{Fishman2022}%
  \BibitemOpen
  \bibfield  {author} {\bibinfo {author} {\bibfnamefont {M.}~\bibnamefont {Fishman}}, \bibinfo {author} {\bibfnamefont {S.~R.}\ \bibnamefont {White}},\ and\ \bibinfo {author} {\bibfnamefont {E.~M.}\ \bibnamefont {Stoudenmire}},\ }\bibfield  {title} {\bibinfo {title} {The {ITensor} {Software} {Library} for {Tensor} {Network} {Calculations}},\ }\bibfield  {journal} {\bibinfo  {journal} {SciPost Physics Codebases}\ }\href {https://doi.org/10.21468/SciPostPhysCodeb.4} {10.21468/SciPostPhysCodeb.4} (\bibinfo {year} {2022}),\ \bibinfo {note} {arXiv:2007.14822 [cs]}\BibitemShut {NoStop}%
\bibitem [{\citenamefont {Haegeman}(2011)}]{Haegeman2011}%
  \BibitemOpen
  \bibfield  {author} {\bibinfo {author} {\bibfnamefont {J.}~\bibnamefont {Haegeman}},\ }\emph {\bibinfo {title} {Variational renormalization group methods for extended quantum systems}},\ \href {http://hdl.handle.net/1854/LU-1908903} {Ph.D. thesis},\ \bibinfo  {school} {Ghent University. Faculty of Sciences} (\bibinfo {year} {2011})\BibitemShut {NoStop}%
\bibitem [{\citenamefont {Pineda}\ \emph {et~al.}(2010)\citenamefont {Pineda}, \citenamefont {Barthel},\ and\ \citenamefont {Eisert}}]{Pineda2010}%
  \BibitemOpen
  \bibfield  {author} {\bibinfo {author} {\bibfnamefont {C.}~\bibnamefont {Pineda}}, \bibinfo {author} {\bibfnamefont {T.}~\bibnamefont {Barthel}},\ and\ \bibinfo {author} {\bibfnamefont {J.}~\bibnamefont {Eisert}},\ }\bibfield  {title} {\bibinfo {title} {Unitary circuits for strongly correlated fermions},\ }\href {https://doi.org/10.1103/PhysRevA.81.050303} {\bibfield  {journal} {\bibinfo  {journal} {Phys. Rev. A}\ }\textbf {\bibinfo {volume} {81}},\ \bibinfo {pages} {050303} (\bibinfo {year} {2010})},\ \bibinfo {note} {publisher: American Physical Society}\BibitemShut {NoStop}%
\bibitem [{\citenamefont {Kraus}\ \emph {et~al.}(2010)\citenamefont {Kraus}, \citenamefont {Schuch}, \citenamefont {Verstraete},\ and\ \citenamefont {Cirac}}]{Kraus2010}%
  \BibitemOpen
  \bibfield  {author} {\bibinfo {author} {\bibfnamefont {C.~V.}\ \bibnamefont {Kraus}}, \bibinfo {author} {\bibfnamefont {N.}~\bibnamefont {Schuch}}, \bibinfo {author} {\bibfnamefont {F.}~\bibnamefont {Verstraete}},\ and\ \bibinfo {author} {\bibfnamefont {J.~I.}\ \bibnamefont {Cirac}},\ }\bibfield  {title} {\bibinfo {title} {Fermionic projected entangled pair states},\ }\href {https://doi.org/10.1103/PhysRevA.81.052338} {\bibfield  {journal} {\bibinfo  {journal} {Phys. Rev. A}\ }\textbf {\bibinfo {volume} {81}},\ \bibinfo {pages} {052338} (\bibinfo {year} {2010})},\ \bibinfo {note} {publisher: American Physical Society}\BibitemShut {NoStop}%
\bibitem [{\citenamefont {Corboz}\ \emph {et~al.}(2010)\citenamefont {Corboz}, \citenamefont {Orús}, \citenamefont {Bauer},\ and\ \citenamefont {Vidal}}]{Corboz2010}%
  \BibitemOpen
  \bibfield  {author} {\bibinfo {author} {\bibfnamefont {P.}~\bibnamefont {Corboz}}, \bibinfo {author} {\bibfnamefont {R.}~\bibnamefont {Orús}}, \bibinfo {author} {\bibfnamefont {B.}~\bibnamefont {Bauer}},\ and\ \bibinfo {author} {\bibfnamefont {G.}~\bibnamefont {Vidal}},\ }\bibfield  {title} {\bibinfo {title} {Simulation of strongly correlated fermions in two spatial dimensions with fermionic projected entangled-pair states},\ }\href {https://doi.org/10.1103/PhysRevB.81.165104} {\bibfield  {journal} {\bibinfo  {journal} {Phys. Rev. B}\ }\textbf {\bibinfo {volume} {81}},\ \bibinfo {pages} {165104} (\bibinfo {year} {2010})},\ \bibinfo {note} {publisher: American Physical Society}\BibitemShut {NoStop}%
\bibitem [{\citenamefont {Efron}(1982)}]{Efron1982}%
  \BibitemOpen
  \bibfield  {author} {\bibinfo {author} {\bibfnamefont {B.}~\bibnamefont {Efron}},\ }\href {https://doi.org/10.1137/1.9781611970319} {\emph {\bibinfo {title} {The Jackknife, the Bootstrap, and Other Resampling Plans}}}\ (\bibinfo  {publisher} {SIAM},\ \bibinfo {year} {1982})\BibitemShut {NoStop}%
\bibitem [{\citenamefont {Luscher}(1986)}]{Luscher1985}%
  \BibitemOpen
  \bibfield  {author} {\bibinfo {author} {\bibfnamefont {M.}~\bibnamefont {Luscher}},\ }\bibfield  {title} {\bibinfo {title} {{Volume Dependence of the Energy Spectrum in Massive Quantum Field Theories. 1. Stable Particle States}},\ }\href {https://doi.org/10.1007/BF01211589} {\bibfield  {journal} {\bibinfo  {journal} {Commun. Math. Phys.}\ }\textbf {\bibinfo {volume} {104}},\ \bibinfo {pages} {177} (\bibinfo {year} {1986})}\BibitemShut {NoStop}%
\end{thebibliography}%
\end{document}